\documentclass[prd,aps,nofootinbib,preprintnumbers]{revtex4}
\usepackage{amsmath,amssymb,epsf}
\begin{document}
\title{Gravity Waves from Tachyonic Preheating after Hybrid Inflation}
\date{\today}
\author{Jean-Fran\c{c}ois Dufaux$^{1}$, Gary Felder$^{2}$, Lev Kofman$^{3}$ and Olga Navros$^{4}$}
\affiliation{$^{1}$ Instituto de F\'isica Te\'orica \ UAM/CSIC,
Universidad Aut\'onoma de Madrid, Cantoblanco, 28049 Madrid, Spain\\
$^{2}$ Department of Physics, Clark Science Center, 
Smith College Northampton, MA 01063, USA\\
$^{3}$ CITA, University of Toronto, 
60 St. George Street, Toronto, ON M5S 3H8, Canada\\
$^{4}$ Department of Mathematics, University of North Carolina Chapel
Hill, CB3250 Philips Hall, Chapel Hill, NC 27599, USA}
\preprint{FTUAM-08-25}
\preprint{IFT-UAM/CSIC-08-90}

\def\lesssim{\mathrel{\hbox{\rlap{\hbox{\lower4pt\hbox{$\sim$}}}\hbox{$<$}}}}
\def\gtrsim{\mathrel{\hbox{\rlap{\hbox{\lower4pt\hbox{$\sim$}}}\hbox{$>$}}}}
\def\Mp{M_{\mathrm{Pl}}}
\def\be{\begin{equation}}
\def\ee{\end{equation}}
\def\bea{\begin{eqnarray}}
\def\eea{\end{eqnarray}}
\newcommand{\picdir}[1]{../#1}

\begin{abstract}
We study the stochastic background of gravitational waves produced from
preheating in hybrid inflation models.  We investigate different
dynamical regimes of preheating in these models and we compute the
resulting gravity wave spectra using analytical estimates and
numerical simulations. We discuss the dependence of the gravity wave
frequencies and amplitudes on the various potential parameters. We
find that large regions of the parameter space leads to gravity waves
that may be observable in upcoming interferometric experiments,
including Advanced LIGO, but this generally requires very small
coupling constants.
\end{abstract}

\maketitle

%%%%%%%%%%%%%%%%%%%%%%%%%%%%%%%%%%%%%%%%%%%%%%%%%%%%%%%%%%%%%%%%%%%%%%%%%

\section{Introduction}

Gravity waves (GW) from the early universe can carry information about
inflation, (p)reheating after inflation, and even about pre-inflation.

During inflation, tensor modes of the classical large scale
cosmological perturbations are produced from the quantum fluctuations
of gravitons \cite{starob79}.  Their amplitude is proportional to the
energy scale of inflation, and their spectrum extends over a wide
range of wavelengths.  Gravitational waves at the cosmological,
long-wavelength part of the spectrum lead to B-mode polarization of
the CMB anisotropy fluctuations.  In high energy models of inflation
(like chaotic inflation), where the ratio of the amplitudes of the
tensor to scalar modes is about $ r \sim 0.1$, the B-mode of
anisotropies should be detectable by forthcoming CMB polarization
experiments.

However, in anticipation of a null signal observation of GW from
inflation, one might still be able to use GW to constrain inflationary
models in ways other than just constraining the overall energy
scale. There are models of inflation where the total number of
inflationary e-folds $N$ exceeds the minimum required to homogenize
the observable universe only by a small margin. For such models with
anisotropic pre-inflationary expansion, gravity waves from
pre-inflation are amplified and may contribute to the large scale CMB
temperature anisotropy \cite{GKP}.  Observational limits on gravity
waves from CMB polarization experiments also result in constraints on
$N$.

If inflation occurs at lower energies (as in many hybrid inflation
models), the amplitude of the resulting gravity waves would be too
weak to be observed with CMB anistropies. However, GW from inflation
at the short-wavelength part of the spectrum might fall in the
amplitude-frequency range which accessible to future gravity wave
astronomy projects such as BBO and DECIGO.  These experiments could
probe $r$ down to the level $r \sim 10^{-6}$.

Gravity waves may also carry unique information about the
post-inflationary dynamics, in particular the inflaton decay and the
subsequent evolution of its decay products towards thermal
equilibrium.  Often  this process starts with preheating, a
violent non-perturbative restructuring of the field configurations
shortly after inflation. Preheating leads to large, non-linear field
inhomogeneities which necessarily generate a classical GW
background. This mechanism of GW production is complementary to the
production of GW from vacuum fluctuations during inflation. In
our previous paper \cite{DBFKU} we developed a machinery for analytic
and numerical calculations of GW production in preheating
models. There are several other papers on the subject that we will
discuss below. It is also worth mentioning that the techniques used to
study GW from preheating have parallels in calculations of GW from
phase transitions / hydrodynamical turbulence, see
e.g. \cite{transitions} and references therein. In \cite{DBFKU} we
developed applications of our methods to the example of preheating
after chaotic inflation. The parameters of these large-field models
are fixed by CMB normalization, which ensures that inflation must
occur at high energy scales.  As a result, the gravity waves produced
from preheating in these models have typical frequencies of order $f
\sim 10^7 - 10^8$ Hz, which is far beyond the observable range (see
Section \ref{Discussion} for observational constraints). As
conjectured in \cite{frag}, preheating after low-energy inflation may
generate GW with lower frequencies $f$. One of the most popular and
most studied models of inflation and preheating is hybrid inflation
\cite{hybrid} with tachyonic preheating \cite{tachpre}. There are many
hybrid inflation models, notably in the context of supergravity (see
\cite{lythriotto} for a review) and string theory (see
\cite{stringinf} for reviews). In these models, the energy scale of
inflation is not fixed by CMB normalization, so it was
optimistically thought that for some parameters the frequency of GW
from preheating may fall into an observable range. The subject of this
paper is GW production from tachyonic preheating, in particular,
assessing how realistic is the range of parameters that may lead to an
observable signal.

Let us briefly review the literature on the subject.  The production
of gravitational waves from preheating was originally studied in
\cite{pregw1} for chaotic inflation, and recently by several groups in
\cite{pregw2, frag, pregw3, juan1, juan2, DBFKU, pregw4, pregw5}.  The
first numerical methods \cite{pregw1, pregw2} were based on the
Weinberg formalism \cite{weinberg}, which, strictly speaking is
applicable only for isolated sources in a Minkowski background.  This
may lead to significant differences in the resulting gravity wave
spectra, as explained in \cite{DBFKU}. In \cite{pregw3}, the gravity
wave equations in an expanding universe were solved in Fourier space.
Another method was used in \cite{juan1, juan2}, where the evolution
equation for metric perturbations were solved in configuration space.
In \cite{juan1} and the earlier version of \cite{juan2} the
transverse-traceless radiative part of GW was not properly
extracted\footnote{Incorrect extraction of TT part, in particular, may
  lead to the incorrect conclusion that a significant amount of
  gravity waves is produced from the stage of scalar field
  ``turbulence'' after preheating, see \cite{DBFKU} for details.}  but
after discussion of the problem in \cite{DBFKU} this was corrected in
the second version of \cite{juan2}.  Finally, in \cite{DBFKU}, we
developed a method based on the Green's function solution in momentum
space, to calculate, numerically and analytically, the production of
gravity waves from a stochastic medium of scalar fields in an
expanding universe. At present, the numerical methods of different
groups \cite{pregw3}, \cite{juan2} and \cite{DBFKU} seem to agree well
with each other, see \cite{pregw4,pregw5} for a comparison of the
results in a model of chaotic inflation.

Most of these papers focused on models of preheating based on
parametric resonance after chaotic inflation, which leads to gravity
wave signals at frequencies which are irrelevant
observationally. Models of preheating after hyrbid inflation are
observationally much more interesting, because they involve extra
parameters and they occur at lower energy scales, and thus they may
generate GW with lower frequencies.  Gravity wave production after
hybrid inflation was first considered in \cite{juangw}, extrapolating
the results of \cite{pregw1}, assuming that preheating occurs through
parametric resonance. However, we know since \cite{tachpre} that
preheating after hybrid inflation occurs in the qualitatively
different regime of tachyonic amplification due to the dynamical
symmetry breaking, when the fields roll towards the minimum through
the region where their effective mass is negative -- a process called
tachyonic preheating.  It completes in one or very few oscillations of the
fields and is very different from parametric resonance\footnote{Hybrid
  inflation was also invoked in \cite{pregw3} to motivate a
  $m^2\,\phi^2 + g^2 \phi^2 \chi^2$ model of parametric resonance at
  low energy scales.  However, as noted above, this model is not
  relevant for preheating after hybrid inflation.  The gravity wave
  spectra shown in \cite{pregw3} fall into an observable range for an
  inflaton mass of order $10$ GeV, instead of $10^{13}$ GeV in chaotic
  inflation (and for a coupling constant of order $10^{-30}$).  One
  can also consider GW production from the non-perturbative decay of a
  scalar field $\phi$ in this model not necessarily related to the
  inflaton, as for example in the scenario considered in
  \cite{curvaton}. The present-day peak frequency $f_*$ may then be
  estimated, in the same way as we do below, from the typical momentum
  $k_* \sim \sqrt{g m \Phi_0} a^{-3/4}$ amplified in this model. If
  $\phi$ dominates the energy density before decaying and the decay is
  followed by the radiation dominated era, one finds $f_* >
  \sqrt{g}\,10^{10}\,\mathrm{Hz}$, independently of the mass $m$ and
  the initial amplitude $\Phi_0$ of the field $\phi$ (as long as $g
  \Phi >> m$ for preheating to occur). This bound can be relaxed if
  $\phi$ does not dominate the energy density, but a very small
  coupling $g^2$ is still required to achieve $f_* <
  10^3\,\mathrm{Hz}$, a necessary condition for these GW to be
  observable.}.

Before further discussion on GW, we have to comment about specific challenges for 
numerical simulations of tachyonic preheating in hybrid inflation models 
with several parameters \footnote{Contrary to the case of resonant preheating after single-parameter
chaotic inflation}. These parameters include the self-coupling $\lambda$ of the symmetry 
breaking fields and their coupling $g^2$ to the inflaton. The scales of the preheating dynamics can be 
significantly  different depending on the combination $g^2/\lambda$ (roughly speaking, the ratio of  
effective masses around the local maximum and minimum of the potential).
Therefore,  care should be taken to incorporate both scales in the simulations.
An additional subtlety is related to the initial conditions for the
fields around the bifurcation point, namely, relatively high or low inflaton velocity,  
which results in qualitatively  different initial regimes -- quantum diffusion or
classical fast or slow rolls around the bifurcation point, see \cite{tachpre} for details. 
The case with higher initial velocity is easier to model numerically.

The production of gravitational waves properly  from tachyonic preheating was first 
investigated in \cite{juan1, juan2}, for a specific region of the parameter 
space ($g^2 \sim \lambda \sim 1$ and significant   velocity of the inflaton at 
the bifurcation point), where the dynamical scales are of the same order.
The resulting gravity wave spectra were located at frequencies too high to be observable, 
as in chaotic inflation models. 
Refs. \cite{juan1, juan2} also displayed a conjecture for a low frequency spectrum, 
but the dependence on the parameters was not studied, so it remains  unclear which model, if any, 
could lead to an observable signal. In \cite{frag}, it was conjectured that 
the peak frequency and amplitude of the gravity wave spectrum depends in a 
relatively simple way on the typical scale amplified during preheating, which 
is usually a known function of the parameters. 
In \cite{DBFKU} we verified this conjecture  numerically 
 for a model of chaotic inflation, and used it to estimate 
analytically the peak of the gravity wave spectrum produced in two different 
models of preheating after hybrid inflation. In particular, we noticed that 
models with low  velocity of the inflaton at the critical point are 
observationally more promising, but it is also more challeging to model this case
with  numerical simulations, as noted above. 

In the present paper, we investigate in detail gravity waves produced
from preheating after hybrid inflation, focusing in particular on
their dependence on various model parameters such as $g^2/\lambda$ and
the field velocities around the bifurcation point.  We study GW
production in several qualitatively different settings of tachyonic
preheating, some of which, including the most interesting cases,
were completely missed in the literature. We use a combination of
analytical estimates and improved numerical simulations to compute the
resulting gravity wave spectra and to identify the regions of the
parameter space which may be relevant for upcoming gravity wave
experiments.

The rest of the paper is organized as follows. In Section \ref{model}, 
we briefly review the hybrid inflation models that we will consider. In Section 
\ref{analytics}, we identify three qualitatively different dynamical regimes of preheating 
in these models and we make analytical estimates of the resulting gravity wave spectra. 
In Section \ref{gwcalculations} we describe  and advance further our numerical method for calculating 
gravity wave production from preheating. We apply this method to preheating after hybrid 
inflation in Section \ref{numerics} and we study the dependence of the gravity 
wave amplitudes and frequencies on the parameters of the model. We then 
discuss which regions of the parameter space may lead to a detectable signal.   
Finally, we discuss the implications of our results and directions for future 
work. Some details of our numerical calculations are given in the appendices.

%%%%%%%%%%%%%%%%%%%%%%%%%%%%%%%%%%%%%%%%%%%%%%%%%%%%%%%%%%%%%%%%%%%%%%%%%%%%%%%%%

\section{Parameters of the Hybrid Inflation Model}
\label{model}

We will consider simple hybrid inflation models with the potential
\begin{equation}
\label{hybrid}
V = {1 \over 4} \lambda \left(\sigma^2 - v^2\right)^2 + {1 \over 2}
g^2 \phi^2 \sigma^2 + V_{inf}(\phi).
\end{equation}
The Higgs field $\sigma$ may in principle be real, complex, or
consisting of any number of components, but the case of a real field
is ruled out because it would lead to the production of dangerous sub-horizon size
domain walls. We have tried simulations with varying numbers of
$\sigma$ components and found the results to be largely
unaffected. The simulations shown below are for a two-component
$\sigma$, for which $\sigma^2$ should be understood as 
$|\sigma|^2 = \sigma_1^2 + \sigma_2^2$ where $\sigma_1$ and $\sigma_2$ 
are two real scalar fields. We take the inflaton $\phi$ to be real for 
simplicity.

\begin{figure}[hbt]
\begin{center}
\begin{tabular}{c|c}
\centering \leavevmode \epsfxsize=8cm
\epsfbox{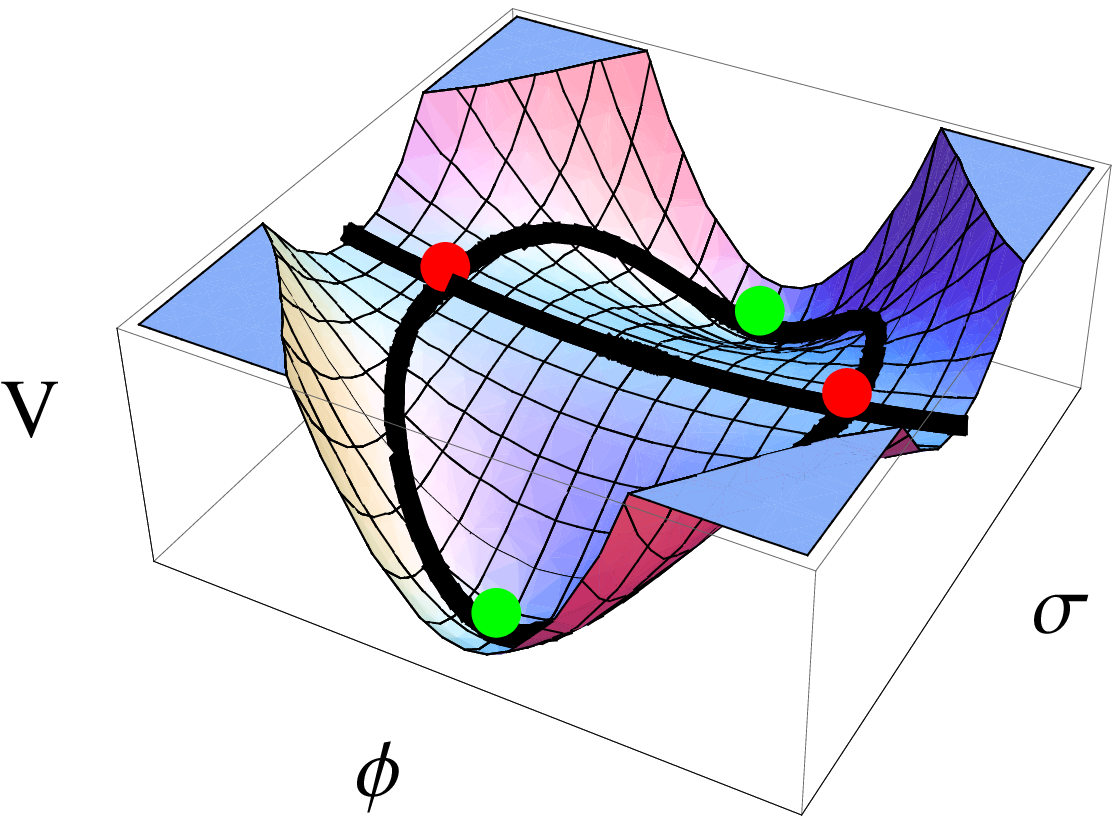} &
\centering \leavevmode \epsfxsize=6cm
\epsfbox{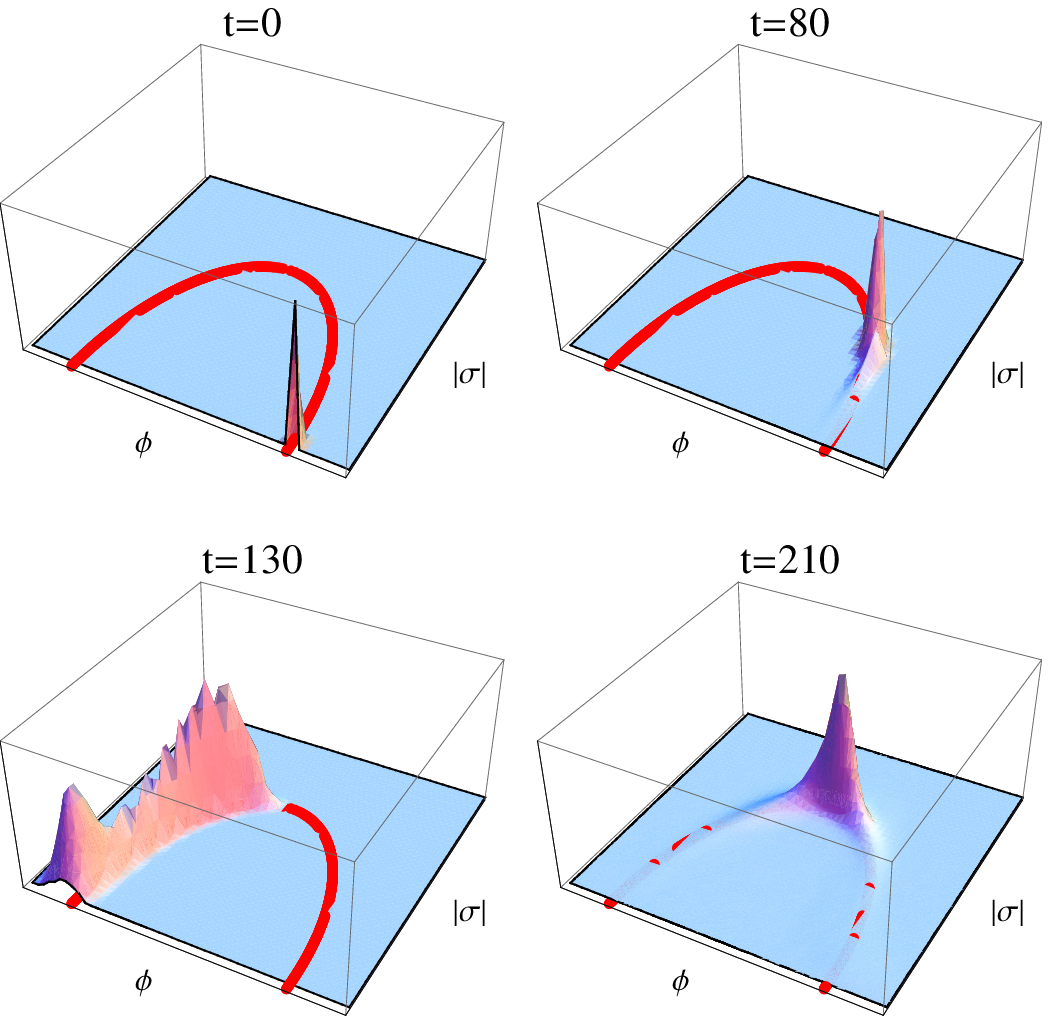}
\end{tabular}
\caption{Hybrid inflation potential. The left panel shows the
  potential as a function of the fields $\phi$ and $\sigma$.  Only one
  direction of $\sigma$ is shown.  The red dots (on the ridge) show
  the critical (bifurcation) points and the green dots (in the
  valleys) show the minima. The inflaton velocity after inflation
  moves it along the top of the ridge. For $g^2 << \lambda$ with a
  small initial velocity the fields roll along the indicated ellipse,
  as can be seen in the field histograms in the right panel.}
\label{hybridpotential}
\end{center}
\end{figure}

For $\phi > \phi_c\,$, where 
\be
\label{defphic}
\phi_c \equiv \frac{\sqrt{\lambda}}{g}\,v \ee is the critical point,
the fields have positive mass squared and the potential has a valley
at $\sigma = 0$. Inflation occurs while $\phi$ decreases slowly in
this valley due to the uplifting term $V_{inf}(\phi)$ in
(\ref{hybrid}). The energy density is usually dominated by the false
vacuum contribution, $V \simeq \lambda v^4 / 4$. Inflation ends either
at the bifurcation point when $\phi = \phi_c$ or when the slow-roll
conditions are violated, whichever occurs first. In both cases, when
$\phi = \phi_c$, $\sigma$ acquires a tachyonic mass and the fields
roll rapidly towards the true minimum at $\phi=0$, $\sigma=v$. During
this rolling field fluctuations are exponentially excited by tachyonic
preheating \cite{tachpre}, thus leading to a rapid decay of the
homogeneous field energy, see Fig.~\ref{hybridpotential}.  This
exponentially rapid growth of inhomogeneities is what drives the
production of gravity waves in this model.

Depending on the model, the slow-roll term $V_{inf}(\phi)$ may take 
various forms, see e.g. \cite{lythriotto}. Usually, it does not 
significantly affect the dynamics during preheating, except by setting 
the  velocity with which $\phi$ reaches the critical point~\footnote{ 
Below we will also often call it the initial velocity at the
onset of preheating, or at the  bifurcation point.}. 
We therefore neglect $V_{inf}$ in the following and consider the initial 
velocity $\dot{\phi}_c$ as a free parameter. We express this velocity in 
dimensionless terms as in \cite{symbreak}
\begin{equation}
\label{defVc}
V_c \equiv \frac{d \tilde{\phi}_c}{d \tilde{t}} = 
\frac{g \, \dot{\phi_c}}{\lambda \, v^2}
\end{equation}
where a dot denotes derivative with respect to the proper time, 
$\tilde{\phi} = \phi / \phi_c$ and $\tilde{t} = m\,t$, 
$m = \sqrt{\lambda} v$ being the natural mass scale for the
potential. The free parameters in the model are thus $\lambda$, 
$g$, $v$, and $V_c$ (not to be confused with the potential $V$).

In cases where inflation lasts until $\phi = \phi_c$, it should end
rapidly enough to avoid the production of Hubble scale inflationary
perturbations of $\sigma$, the so-called waterfall condition
\cite{hybrid, primBH, juanlinde}. Often, the whole process of
preheating occurs in less than a Hubble time. Therefore, in the
following, we will neglect the expansion of the universe. (This is
always accurate for sufficiently small $v$). In this case, defining
new field and spacetime variables $\phi_{new} \equiv \phi/v$,
$\sigma_{new} \equiv \sigma/v$, $x_{\mu,new} \equiv v x_{\mu}$ the
value of $v$ drops out of the field equations of motion and the
amplitude of the initial vacuum fluctuations.  Thus the $v$-dependence
of physical quantities is known analytically and the $v$-dependence of
the gravity wave spectrum displayed below is exact. Specifically,
changing $v$ does not shift the frequencies at all and changes the
density in gravity waves today proportionally to $v^2$.

Depending on the remaining parameters $\lambda$, $g$ and $V_c$, the
way the tachyonic instability develops in the model (\ref{hybrid}) may
occur in different regimes, as we will discuss in the next Section.

%%%%%%%%%%%%%%%%%%%%%%%%%%%%%%%%%%%%%%%%%%%%%%%%%%%%%%%%%%%%%%%%%%%%%%%%%%%%%%%

\section{Analytical Estimates of GW Frequencies and Amplitudes for Different Regimes of Tachyonic Preheating}
\label{analytics}

Before turning to the numerical calculations of GW in the following Sections, 
we estimate analytically how the GW spectra from tachyonic preheating vary with the model 
parameters.

Preheating in the model (\ref{hybrid}) starts when $\phi =
\phi_c$. Because of its initial velocity, the inflaton rolls
classically away from $\phi_c$. At the same time, the Higgs field
acquires a negative mass squared which increases with time starting
from zero at the critical point. This allows for a tachyonic
amplification of its initial quantum fluctuations. When these
fluctuations become comparable to $v$, the field distributions may
settle rapidly around the true minimum at $\phi = 0$, $\sigma =
v$. The preheating process in this case has been studied in detail in
\cite{symbreak}, see also \cite{Copeland:2002ku}. We will briefly
review this case, where preheating is driven by the inflaton initial
velocity, in sub-section \ref{highvinit}. On the other hand, if the
initial velocity of the inflaton is sufficiently low, the classical
rolling of the inflaton may be subdominant and preheating may start in
a different way. The initial quantum fluctuations themselves may
induce a negative curvature of the potential around the critical
point. The tachyonic amplification is then triggered by the quantum
fluctuations instead of the inflaton's classical rolling. We will
discuss this case in sub-section \ref{lowvinit}, and estimate the
initial velocity at which the dynamics crosses between these two
regimes. In both cases of low and high initial velocity, when the
Higgs fluctuations become of the order of the symmetry breaking scale,
the field distributions settle rapidly around the true minimum if $g^2
\gtrsim \lambda$. On the other hand, for $g^2 << \lambda$, a significant fraction of the
energy density is still in the relatively homogeneous inflaton, which
oscillates more than once  around $\phi = 0$ with large amplitude. As we
will see in sub-section \ref{g2lll}, this leads to interesting
differences in the process of GW production.

Of particular interest to us will be the characteristic physical momentum 
$k_*$ of the scalar fields amplified in different regimes of tachyonic preheating.
Indeed, in \cite{frag}   the frequency and 
amplitude of the produced GW were connected to the
dynamics of the ``bubbly'' inhomogeneities associated with the peaks of the random gaussian 
field of fluctuations amplified by preheating. They 
 depend in a simple way on the typical size $R_* \sim 1/k_*$ of 
the field bubbles.  We verified numerically in \cite{DBFKU}, for a model 
of chaotic inflation, that the main contribution to gravity wave production during preheating 
comes indeed from the violent ``bubbly stage'' between the linear and turbulent stages, with 
peak frequency and  amplitude (in terms of the present-day GW energy density)  given by
\begin{eqnarray}
\label{fR}
f_* &\approx& {4 \times 10^{10} Hz \over R_* \rho_p^{1/4}} \ ,\\
\label{omegaR}
h^2 \Omega^*_{gw} &\approx& \alpha \times \,10^{-5}\, (R_* H_p)^2 \ ,
\end{eqnarray}
where $H_p$ and $\rho_p$ are the Hubble parameter and the total energy
density at preheating when gravity waves are produced. Here we have
included a ``fudge factor'' $\alpha$ which may vary from one model to
another ($\alpha \approx 0.1$ for the $\lambda \phi^4$ model
considered in \cite{DBFKU}).  The factor $10^{-5}$ arises from the
redshift of the GW radiation.  Not coincidentally a formula similar to
(\ref{fR}) arises in the theory of GW production from the first order
phase transition with the nucleation of bubbles of the size $R$.

In many models of preheating, it is possible to estimate $k_*$, and therefore according 
to (\ref{fR}) and (\ref{omegaR}), the peak frequency and amplitude of the GW spectrum, as a function of the 
parameters. We do that below for the model (\ref{hybrid}). In this case, when expansion of the universe 
is negligible, we have
\begin{eqnarray}
\label{fkstar}
f_* &\sim& {k_* \over \lambda^{1/4} v}\; 6 \times 10^{10} Hz \ ,\\
\label{Omegakstar}
h^2 \Omega^*_{gw} &\sim& 2 \times 10^{-6} \; {\lambda v^4 \over k_*^2 \, \Mp^2} \ ,
\end{eqnarray}
where we have used $\rho_p = \lambda v^4 / 4$ and $H_p^2 = 8\pi \rho_p / (3 \Mp^2)$. At this level 
we take $\alpha \approx 0.1$ and consider simple analytical estimates. The precise dependence on the 
parameters will be studied numerically in Section \ref{numerics}.

Next, we will estimate the scale $k_*$. It turns out that this depends, in particular, on the onset of preheating.
The dynamics of the fields around the bifurcation (critical)  point is dominated either by the
classical, inertial motion of the field $\phi$ superposed with the  quantum fluctuations of $\sigma$, or by
quantum fluctuaions (quantum diffusion) of both fields. The estimates will depend on which process is dominant.

%%%%%%%%%%%%%%%%%%%%%%%%%%%%%%%%%%%%%%%%%%%%%%%%%%%%%%%%%%%%%%%%%%%%%%%%

\subsection{Onset of Tachyonic Preheating  by the Rolling Inflaton}
\label{highvinit}

Because of its initial velocity at the critical point, the inflaton
field $\phi$ rolls classically away from $\phi_c$ along the ridge of
the potential at $\sigma = 0$.  As a result, the Higgs field acquires
a tachyonic mass $- m_{\sigma}^2 \,=\, \lambda v^2 - g^2 \phi^2 \,=\,
g^2 (\phi_c^2 - \phi^2) \, $, which increases in time starting from
$m_{\sigma}^2 = 0$ at the critical point. This allows for an
exponential amplification of the quantum fluctuations of $\sigma$ with
momenta $k^2 < - m_{\sigma}^2$. If the inflaton has sufficient
velocity (see below), this process dominates the beginning of
preheating.

In this case, we can estimate the typical momenta amplified initially
as follows \cite{juanlinde}, see also \cite{Copeland:2002ku,
  symbreak}. Close to the critical point, the tachyonic mass of
$\sigma$ increases as $- m_{\sigma}^2 \,=\, g^2 (\phi_c^2 - \phi^2)
\,\simeq\, 2g^2\,\phi_c\,|\dot{\phi}_c|\,\Delta t \,$ after a time
$\Delta t$ from the moment when $\phi = \phi_c\,$. The resulting
exponential growth of quantum fluctuations becomes efficient when
$\Delta t \gtrsim \sqrt{- m_{\sigma}^2} \, $, leading to a typical
momentum $k_*^2 \lesssim - m_{\sigma}^2 \,$ given by \be
\label{kstarV}
k_*^3 \approx 2g^2\,\phi_c\,|\dot{\phi}_c| = 2 V_c \, m^3 \ee where we
have used (\ref{defVc}). A broader range of momenta will be amplified
as $- m_{\sigma}^2$ continues to increase, but the modes with lower
momentum (\ref{kstarV}) will already have an exponentially higher
amplitude and their subsequent growth will occur exponentially
faster. The process should thus be dominated by the modes with typical
momentum given by (\ref{kstarV}).

Inserting this result into Eqs.~(\ref{fkstar}), (\ref{Omegakstar}) gives
\begin{eqnarray}
\label{fstarVc}
f_* &\sim& \lambda^{1/4} \, V_c^{1/3} \; 7 \times 10^{10} Hz \ , \\
\label{OmegastarVc}
h^2 \Omega^*_{gw} &\sim& 10^{-6} \, V_c^{-2/3} \, \left(\frac{v}{\Mp}\right)^2 \ .
\end{eqnarray}

We have checked numerically that for a broad range of parameters the typical momenta amplified at 
the \emph{beginning} of preheating are in very good agreement with Eq.~(\ref{kstarV}) in the case 
of significant intial velocity. However, for $g^2 / \lambda << 1$, the typical momenta are significantly 
shifted towards the infra-red \emph{after} the first tachyonic growth. We will discuss this case 
separately below. We will see that, except in this case, our numerical results for the gravity wave 
spectrum are well described by Eqs.~(\ref{fstarVc}), (\ref{OmegastarVc}). They show that, for a given 
non-negligible initial velocity $V_c$, the peak frequency depends on the energy density $\lambda v^4$ 
only through $\lambda^{1/4}$, while the peak amplitude varies as $v^2$, see \cite{DBFKU}. They also show 
that the smaller the initial velocity $V_c$, the lower the frequency and the higher the amplitude. 
However, these formulas cannot be extrapolated to arbitrary low initial velocity of the inflaton, as we will now discuss.

%%%%%%%%%%%%%%%%%%%%%%%%%%%%%%%%%%%%%%%%%%%%%%

\subsection{Quantum Diffusion Onset of Tachyonic  Preheating}
\label{lowvinit}

We discuss below the conditions that should apply to consider the classical inflaton rolling to be subdominant,
so to simplify the discussion let us first consider a vanishing initial velocity at the critical point.

In this case, reasoning as above, one would conclude that 
the fields stay relatively long at the critical point $\phi = \phi_c\,$, $\sigma = 0\,$, 
since the curvature of the potential vanishes there. 
However, quantum fluctuations of the fields around the critical point 
move the fields to the region of negative curvature slope of the potential.
 The resulting instability may be 
described as quantum diffusion away from the critical point of some modes among the 
initial fluctuations, due to their interactions with the other modes \cite{tachpre}. 

In order to estimate the typical momentum amplified by this process, we first 
determine the direction in field space along which the fields are most likely to 
roll from the rest, i.e. the directions of steepest potential in the vicinity of the critical point. 
Letting $\sigma = r \, \mathrm{sin}\theta$ and $\phi = \phi_c - r \, \mathrm{cos}\theta \,$, 
the potential reduces to
\be
V(r, \theta) = \frac{\lambda}{4} \, v^4 - 
g \, \sqrt{\lambda} \, v \, r^3 \, \mathrm{cos}\theta \, \mathrm{sin}^2\theta + \mathcal{O}(r^4) 
\ee
at small distance $r$ (in the fields space) from the critical point. At fixed $r$, this potential is 
minimum for $\mathrm{tan}\theta = \pm \sqrt{2} \,$. The effective potential along these 
directions is given by
\be
\label{Veffr}
V(r) = \frac{\lambda}{4} \, v^4 - \frac{2}{3\sqrt{3}} \, 
g \, \sqrt{\lambda} \, v \, r^3 + \mathcal{O}(r^4) \, .
\ee

Tachyonic preheating for such a cubic potential has been considered in
\cite{tachpre}.  Let us briefly review their results. The (canonically
normalised) scalar field $r$ has initial quantum fluctuations with
amplitude $|\delta r_k| \,\sim\, 1 / \sqrt{2k}$ around the critical
point. Consider long-wavelength modes with $k \,\lesssim\, k_0$, for
some cut-off $k_0$. Their contribution to the mean-square fluctuations
is given by $\langle r^2(k_0) \rangle \,=\, (2\pi)^{-3} \,
\int_0^{k_0} d^3k \, |\delta r_k|^2 \,\sim\, k_0^2 / (8\pi^2)\,$.
Thus short-wavelength fluctuations, with momenta $k = \gamma k_0$ for
$\gamma$ somewhat greater than one, may be seen to live on top of a
quasi-homogeneous long-wavelength field $r$ with an average amplitude
$r \,\sim\, r_{\mathrm{rms}}(k_0) \,\sim\, k_0 / (2\sqrt{2} \pi) \,
$. These short-wavelength fluctuations feel a negative curvature
induced by the long-wavelength field $r$: $V''(r_{\mathrm{rms}}(k_0))
\,=\, -4\,g\,\sqrt{\lambda}\,v\,r_{\mathrm{rms}}(k_0) / \sqrt{3}\,$.
This may lead to a tachyonic amplification of the short-wavelength
modes with momenta $k^2 \,\lesssim\, |V''| \,\sim\, \sqrt{2
  \lambda}\,g\,v\,k_0 / \sqrt{3 \pi^2} \,$. Taking for definiteness
$\gamma \gtrsim \sqrt{2}\,$, one may argue that fluctuations with $k
\,\lesssim\, g\,\sqrt{\lambda}\,v / \sqrt{3 \pi^2}$ may enter a
self-sustained regime of tachyonic growth. A more careful
investigation \cite{tachpre} shows that modes with somewhat higher
momenta, even if initially more suppressed, grow faster and tend
therefore to dominate the instability. We will write the
characteristic momentum amplified by this process as \be
\label{kstarV0}
k_* \approx C\,g\,\sqrt{\lambda}\,v
\ee
where $C$ is a numerical constant to be determined emprirically. 

The estimates above neglect the inflaton initial velocity. 
These estimates are presumably accurate when the corresponding tachyonic growth 
occurs faster than the one due to the classical rolling of the inflaton. 
Roughly speaking, that will be true when the typical momentum amplified is larger  
than the typical momentum amplified by the classical rolling. 
We thus expect the estimates above to be valid when 
\be
\label{condVnegl}
k_{*\,\mathrm{quant}} > k_{*\,\mathrm{class}} \;\;\;\;\;\; \Leftrightarrow 
\;\;\;\;\;\; V_c < \frac{C^3\,g^3}{2} \;\;\;\;\;\; \Leftrightarrow 
\;\;\;\;\;\; \dot{\phi}_c < \frac{C^3}{2}\,g^2\,\lambda\,v^2 \ ,
\ee
where we have used (\ref{kstarV0}), (\ref{kstarV}) and (\ref{defVc}).

Inserting (\ref{kstarV0}) into Eqs.~(\ref{fkstar}), (\ref{Omegakstar}) gives
\begin{eqnarray}
\label{fstarV0}
f_* &\sim& C\,g\,\lambda^{1/4}\; 6 \times 10^{10} Hz \ ,\\
\label{OmegastarV0}
h^2 \Omega^*_{gw} &\sim& \frac{2 \times 10^{-6}}{C^2\,g^2} 
\,\left(\frac{v}{\Mp}\right)^2 \, .
\end{eqnarray}
Comparing to the case of significant initial velocity (\ref{fstarVc}), we see that it is 
much easier to lower the gravity wave frequencies using small coupling constants in the case 
of negligible velocity $V_c$.
For $g^2 = 2 \lambda\,$, we recover the results $f_* \propto \lambda^{3/4}$ and 
$h^2 \Omega^*_{gw} \propto v^2 / \lambda$ of \cite{DBFKU}.

%%%%%%%%%%%%%%%%%%%%%%%%%%%%%%%%%%%%%%%%%%%%%%%%%%%%%%%%%%%%%%%

\subsection{Successive GW production  for $g^2 << \lambda$}
\label{g2lll}

We found that a qualitatively different regime of GW production from tachyonic preheating in hybrid inflation 
occurs for  $g^2 << \lambda$. In this case, preheating starts as in the previous sub-sections, 
but when the Higgs fluctuations become of the order of the symmetry breaking scale, 
$\delta \sigma \sim v$, the inflaton is still relatively homogeneous, $\delta \phi << \phi_c$, 
and makes more than one oscillation   with large amplitude around $\phi = 0$. 

Indeed, the effective mass 
of the inflaton, $m^2_\phi \,=\, g^2\,\langle\sigma^2\rangle$ is much smaller in this case, 
see also \cite{juanlinde}. As a result, we will see that the characteristic momentum which 
has been amplified \emph{at the end} of preheating may differ significantly from (\ref{kstarV}) 
or (\ref{kstarV0}). 

\begin{figure}[hbt]
\begin{center}
\begin{tabular}{ccc}
\centering \leavevmode \epsfxsize=6cm
\epsfbox{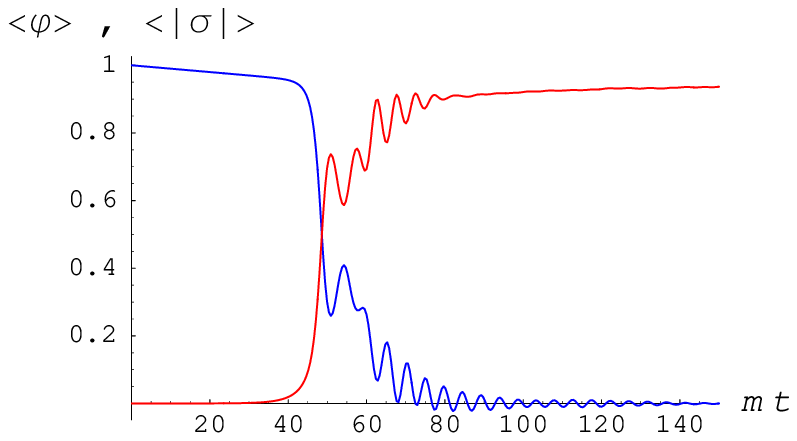} &
\centering \leavevmode \epsfxsize=6cm
\epsfbox{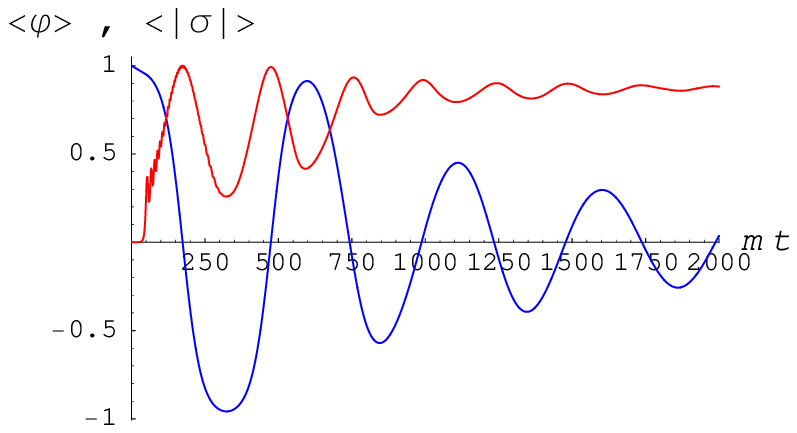} &
\centering \leavevmode \epsfxsize=6cm
\epsfbox{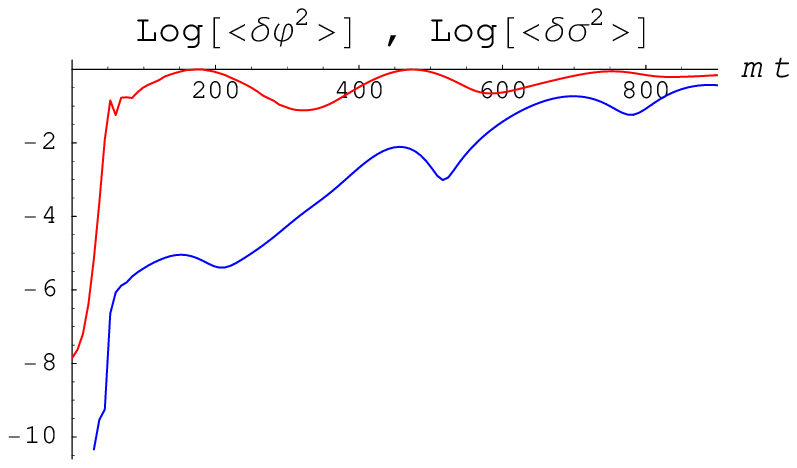}
\end{tabular}
\caption{The left and middle panels show the evolution with time of the inflaton's 
mean normalized to $\phi_c$ (blue), $\langle\phi / \phi_c\rangle$, and of the mean 
of the Higgs modulus normalized to its vev (red), $\langle|\sigma|/v\rangle$ 
(where $|\sigma|^2 = \sigma_1^2 + \sigma_2^2$), for $\lambda / g^2 = 0.5$ (left panel) 
and $\lambda / g^2 = 2000$ (middle panel). The other parameters are $V_c = 10^{-3}$ 
and $\lambda = 10^{-5}$ in both cases. The right panel shows the evolution with time 
of the variances $\mathrm{Log}[\langle\delta \phi^2 / \phi_c^2\rangle]$ (blue) and 
$\mathrm{Log}[\langle\delta \sigma^2 / v^2\rangle]$ (red) for the case 
$\lambda / g^2 = 2000\,$.}
\label{meanslg}
\end{center}
\end{figure}

Fig.~\ref{meanslg} shows the evolution with time of the averages $\langle\phi\rangle$ and 
$\langle|\sigma|\rangle$ (in the case of significant  velocity at the preheating onset),  
for $g^2 \sim \lambda$ (left panel) and $g^2 << \lambda$ (middle panel), 
and the same values of the other parameters. The right panel shows the evolution with 
time of the variances of the fields in the case $g^2 << \lambda$. One sees clearly the large 
inflaton oscillations around the true minimum, roughly from $\phi_c$ to $-\phi_c$ for the 
first oscillations in the case of the middle panel of Fig.~\ref{meanslg}. Note in particular 
that each time the inflaton approaches $\pm \phi_c$, the minimum in the $\sigma$-direction 
is at $\sigma = 0$, so that $\sigma$ rolls back towards the origin where the symmetry is 
restored. 

In both cases of Fig.~\ref{meanslg}, $\langle|\sigma|\rangle$ is first rapidly 
amplified up to a value a few times smaller than its VEV by the tachyonic 
effect as $\phi$ rolls slowly away from $\phi_c$ due to its initial velocity. 
The typical momentum amplified by this process is given by (\ref{kstarV}) and 
is independent of $g^2 / \lambda$. After that stage, non-linearities become 
significant. Because of the amplification of $\sigma$, $\phi$ acquires an 
effective mass which makes it roll faster towards $\phi = 0$. This happens 
much more slowly for $g^2 << \lambda$ since $m_\phi$ is much smaller in 
that case. As $\phi$ decreases, the tachyonic mass of $\sigma$ momentarily increases, which 
amplifies further $\sigma$ fluctuations until backreaction shuts off the tachyonic 
effect. For $g^2 \gtrsim \lambda$, the field distributions then rapidly settle around 
the true minimum, with dispersions $|\delta \sigma| \sim v$ and $|\delta \phi| \sim \phi_c$. 
On the other hand, for $g^2 << \lambda$, we still have $|\delta \phi| << \phi_c$, see the 
right panel of Fig.~\ref{meanslg}.

We can further study this stage by noting that the trajectory of the
field distributions in field space is quite accurately given for some
time (up to $m t$ of order $600$ in the case $\lambda / g^2 = 2000$ of
the middle panel of Fig.~\ref{meanslg}) by the ellipse \be
\lambda\,\sigma^2 + g^2\,\phi^2 = \lambda\,v^2 \; , \ee which is
essentially the condition $\partial V / \partial\sigma = 0$. This is
satisfied when $|\phi| < \phi_c$, see Fig.~\ref{hybridpotential} for illustration.
 Along this trajectory, the potential reduces to
\be
\label{effV}
V = \frac{g^2}{2}\,v^2\,\phi^2 - \frac{g^4}{4\,\lambda}\,\phi^4 \ ,
\ee
and the dynamics corresponds to a single-field system with effective mass
\be
\frac{m^2_\phi}{m^2} = \frac{g^2}{\lambda}\,
\left(1-3\,\frac{\phi^2}{\phi_c^2}\right) \ ,
\ee
where $\langle\phi^2\rangle \, \simeq \, \langle\phi\rangle^2$ at that stage. 
We see that the mass squared becomes negative each time that $|\phi| > \phi_c / \sqrt{3}$. 
This leads to the tachyonic growth of modes with typical momenta 
\be
\label{kstarg}
k_* \sim g\,v\,.
\ee 
Note that this is much smaller than (\ref{kstarV}) for $g^2 << \lambda$. 
One can clearly see these successive tachyonic growths after the first, much more rapid 
amplification, in the right panel of Fig.~\ref{meanslg}.
The successive tachyonic growths and the resulting bubble inhomogeneities will lead to succesive bursts
of GW productions. We will confirm this effect numerically in Section VC.

In addition to these tachyonic growths, there are other non-adiabatic amplifications 
when $m_{\phi}(\phi) \approx 0$, which significantly affect the spectrum of $\phi$ 
fluctuations. This is very similar to the preheating process in models of 
new inflation \cite{Desroche:2005yt}. Indeed, for the potential (\ref{effV}) the 
non-adiabaticity condition on the frequency, $|\dot{\omega}| \gtrsim \omega^2$, 
reads $\left(k^2 + m^2_\phi\right)^{3/2} \lesssim 3\,|\phi|\,|\dot{\phi}|\,g^4 / \lambda$. 
This is most easily satisfied around $\phi \approx \pm \phi_c / \sqrt{3}$, where 
$m_{\phi} \approx 0$. At that point, $|\dot{\phi}|$ may in principle depend 
on the initial inflaton velocity during the first oscillations, but
for the last few
non-adiabatic amplifications a good approximation is given by 
$|\dot{\phi}| \sim \sqrt{2 \lambda}\,v^2 / 3$. This leads to the growth of modes 
with $k^2 \lesssim g\,v$ around $\phi \approx \pm \phi_c / \sqrt{3}$. The typical 
momentum is lower by a factor of $2$ or so, $k_* \approx g\,v / 2$.

Inserting this result into Eqs.~(\ref{fkstar}), (\ref{Omegakstar}) gives
\begin{eqnarray}
\label{fstarg}
f_* &\sim& \frac{g}{\sqrt{\lambda}}\,\lambda^{1/4}\; 3 \times 10^{10} Hz \ ,\\
\label{Omegastarg}
h^2 \Omega^*_{gw} &\sim&  8 \times 10^{-6}\,\frac{\lambda}{g^2}\,
\left(\frac{v}{\Mp}\right)^2 \, .
\end{eqnarray}
Note that this does not depend anymore on the initial velocity $V_c$.
Compared to (\ref{fstarVc}, \ref{OmegastarVc}), we see that usually the peak 
frequency decreases and the peak amplitude increases for $g^2 << \lambda\,$. 

Since Eqs.~(\ref{fstarg}), (\ref{kstarg}) do not depend on the initial tachyonic 
amplification, we expect them to hold also in the case of negligible initial 
velocity. Note that, in this case, it is more difficult than in (\ref{fstarV0}) 
to lower $f_*$ with small coupling constants. Thus it is observationally more 
interesting to have $g^2 \gtrsim \lambda$ in the case of negligible initial velocity. 

%%%%%%%%%%%%%%%%%%%%%%%%%%%%%%%%%%%%%%%%%%%%%%%%%%%%%%%%%%%%%%%x

\section{Calculation of Gravity Waves}
\label{gwcalculations}

In this Section we refine the  basic method of GW calculations that we 
developed in \cite{DBFKU}.

The energy density of gravity waves is constructed from the amplitude
of the transverse-traceless part of the metric perturbation
$h_{ij}$. The TT part can be extracted by applying a momentum-space
projection operator $O_{ijlm}(k)$ (see e.g. \cite{DBFKU}), but this
requires calculating the metric perturbations in momentum
space. Alternatively, following \cite{juan2}, we can solve the
position-space evolution equations for the whole $h_{ij}$ and then,
when calculating gravity wave spectra, apply the projection operator
to the end-point of the evolution results. Since the equations of
motion and the projection operator are both linear they commute.

The equation of motion for the metric perturbations is
\begin{equation}\label{eom}
\bar{h}_{ij}'' - \nabla^2 \bar{h}_{ij} - {a'' \over a} \bar{h}_{ij} =
16 \pi G a^3 \Pi_{ij}^{TT},
\end{equation}
where we are using conformal time $d\tau = dt/a$ and metric
perturbations $\bar{h}_{ij}=a h_{ij}$. We will use the non-TT source
term
\begin{equation}
\Pi_{ij} = {1 \over a^2} \partial_i \phi \partial_j \phi \ ,
\end{equation}
and compensate by applying the projection operator (in momentum space)
to the resulting $\bar{h}_{ij}$. The other terms in the
energy-momentum tensor  vanish under the
TT projection. The results of this calculation read as 
\begin{equation}
\label{green1}
\bar{h}_{ij}(\tau, \mathbf{k}) = \frac{16 \pi
  G}{k}\,\int_{\tau_i}^{\tau} d\tau'\ \, {\cal G}\left[k\,(\tau, \tau')\right]\, a(\tau')\,T_{ij}^{\mathrm{TT}}(\tau', \mathbf{k}) \ ,
\end{equation}
where $ {\cal G}$ is the Green's function of the operator
$\partial^2_t+k^2-{a'' \over a}$. In the typical case of power-law growth of the scale factor, 
$a(t) \sim t^{\alpha}$, 
the Green's function is easily constructed from Bessel functions,
and may have different behaviours for the sub- and super-horizon modes.
During preheating, the equation of state $w$ usually jumps
rapidly to a value close to $1/3$ \cite{eos}.
In a radiation-dominated universe we can put ${a'' \over a}=0$. In this case the general
 solution (\ref{green1}) acquires a simple form
\begin{equation}
\label{green}
\bar{h}_{ij}(\tau, \mathbf{k}) = \frac{16 \pi
  G}{k}\,\int_{\tau_i}^{\tau} d\tau'\ \,\sin\left[k\,(\tau -
  \tau')\right]\, a(\tau')\,T_{ij}^{\mathrm{TT}}(\tau', \mathbf{k}) \ ,
\end{equation}
which can be matched to the source-free solution to give, at late
times
\begin{equation}\label{hfree}
\bar{h}_{ij}(\tau, \mathbf{k}) = A_{ij}(\mathbf{k})\,\sin\left[k (\tau - \tau_f\
)\right] +
B_{ij}(\mathbf{k})\,\cos\left[k (\tau - \tau_f)\right] \;\;\;\; \mbox{ for } \;\
\;\; \tau \geq \tau_f,
\end{equation}
where
\begin{eqnarray}
A_{ij}(\mathbf{k}) &=& \frac{16\pi G}{k}\,\int_{\tau_i}^{\tau_f} d\tau'\,\cos\left[k\,(\tau_f - \tau')\right]\,
a(\tau')\,T_{ij}^{\mathrm{TT}}(\tau', \mathbf{k})
\nonumber \\
B_{ij}(\mathbf{k}) &=& \frac{16\pi G}{k}\,\int_{\tau_i}^{\tau_f} d\tau'\,\sin\left[k\,(\tau_f - \tau')\right]\,
a(\tau')\,T_{ij}^{\mathrm{TT}}(\tau', \mathbf{k}) \ .
\end{eqnarray}

Once we have Fourier transformed the metric perturbations and
projected out the TT part the gravity wave energy density can be calculated as
\begin{equation}
\rho_{gw} = {1 \over 32 \pi G a^4} {1 \over V} \int d^3k
\bar{h}_{ij}'(\vec{k}) \bar{h}_{ij}'^*(\vec{k}) \ ,
\end{equation}
where summation over $i$ and $j$ is understood. 
However, this formula will include small time oscillations of the modes, given by
the sine and cosine terms in Eq.~(\ref{hfree}) above. In our Fourier
space calculations we eliminated these oscillations by averaging over
a full period, thus replacing $\bar{h}_{ij}'(k) \bar{h}_{ij}'^*(k)$
with $(k^2/2)\left(|A_{ij}|^2 + |B_{ij}|^2\right)$. We can accomplish
the same thing in this calculation by noting that at time
$\tau=\tau_f$ we have $B_{ij} = \bar{h}_{ij}$, $A_{ij} =
\bar{h}_{ij}'/k$, so we can calculate the energy density averaged over
a full oscillation as
\begin{equation}\label{energy3}
\rho_{gw} = {1 \over 32 \pi G a^4} {1 \over V} \int d^3k {k^2 \over 2}
\left(|\bar{h}_{ij}|^2 + {1 \over k^2} |\bar{h}_{ij}'|^2\right) = {1
  \over 64 \pi G a^4} {1 \over V} \int d^3k \left(k^2 |\bar{h}_{ij}|^2
+ |\bar{h}_{ij}'|^2\right) \ .
\end{equation}
We use the notation $|X_{ij}|^2 = \sum_{i,j} X_{ij} X_{ij}^*$.

By assuming isotropy we can write
\begin{equation}
{1 \over 16 G a^4} {1 \over V} \int dk k^2 \left(k^2 |\bar{h}_{ij}|^2
+ |\bar{h}_{ij}'|^2\right) \ .
\end{equation}
The gravity wave spectrum can then be calculated as
\begin{equation}
\label{Sk}
S_k \equiv a^4 k {d \rho_{gw} \over dk} = {k^3 \over 16 G V} \left(k^2
|\bar{h}_{ij}|^2 + |\bar{h}_{ij}'|^2\right) \ .
\end{equation}

We make a couple of remarks about the formulas (\ref{energy3})-(\ref{Sk}).
As we demonstrated in \cite{DBFKU}, in the limit of $\tau_f \to \infty$ and
$a(t) \to const$, we recover Weinberg's formula for the
emission of GW from isolated sources in flat space-time. 
Next, the forms (\ref{energy3})-(\ref{Sk}) admit a natural interpretation in terms
of a number density of emitted gravitons.
Indeed, from the oscillating amplitudes $\bar{h}_{ij}(\vec k, \tau)$ one can construct
the adiabatic invariant
\begin{equation}\label{number}
n_k=\frac{1}{2k} \left(k^2 |\bar{h}_{ij}|^2 + |\bar{h}_{ij}'|^2\right) \ ,
\end{equation}
which corresponds to the number of gravitons per mode $\vec k$.
Then, Eq.~(\ref{energy3}) can be re-interpreted in terms of the
energy density carried out by the gravitons
\begin{equation}\label{energy4}
\rho_{gw} =  {1 \over 32\pi G a^4} {1 \over V} \int d^3k \, k \, n_k \ .
\end{equation}
Correspondingly, $S_k={k^4 n_k \over 8 G V}$.

From Eq.~(\ref{Sk}), the derivation of today's spectrum occurs as it did with
our Fourier space calculation, adapted for the hybrid inflation case
as discussed below. In the chaotic inflation case we evolved the
spectrum to today by defining a time $t_j$ at the end of the
simulation after which we took the equation of state to be $w=1/3$. In
the case of hybrid inflation, we are not considering expansion during preheating and we
assume that the equation of state approaches $w=1/3$ within a few
Hubble times afterwards, so we effectively take $t_j$ to be the end of
inflation, and simply evolve to today's variables with
\begin{equation}
f = {k \over \rho^{1/4}} (4 \times 10^{10} Hz) \ ,
\end{equation}
\begin{equation}
\label{omega}
\Omega_{gw} h^2 = (9.3 \times 10^{-6}) {S_k(\tau_f) \over \rho} \ ,
\end{equation}
where we take $\rho = (1/4) \lambda v^4$.

In the next Section we will describe results of numerical calculations of GW 
from lattice simulations of tachyonic preheating.
However, the infra-red (IR) tail of the spectrum, which is often interesting for observations,
is not captured by the finite-size simulations. 
Eq.~(\ref{green}) is useful for deriving the IR asymptotics 
of the gravity wave spectrum. For momenta small compared to the peak of 
the scalar field spectra, $k << k_*$, the convolution of the scalar 
field spatial derivatives is independent of $\mathbf{k}$, 
$T_{ij}^{\mathrm{TT}}(\tau', \mathbf{k}) \simeq T_{ij}^{\mathrm{TT}}(\tau', 0)$. 
This corresponds to the quadrupole approximation. We can then distinguish 
two different regimes, depending on the time variation of the Green's function 
$\sin\left[k\,(\tau -\tau')\right]$ and the source $T_{ij}^{\mathrm{TT}}(\tau', 0)$. 
For sufficiently small $k$, the Green's function varies more slowly than the source  
and may be taken out of the integral in (\ref{green}). This gives 
$\Omega_{gw} \propto f^3$ for the IR tail of the GW spectrum (\ref{omega}), (\ref{Sk}).
 On the other hand, 
there can be an intermediate frequency range between the IR tail and the peak 
where $\sin\left[k\,(\tau -\tau')\right]$ varies more rapidly than 
$T_{ij}^{\mathrm{TT}}(\tau', 0)$. The time integral in (\ref{green}) then gives 
an extra $1/k$ factor, leading to $\Omega_{gw} \propto f$ for some frequency 
range below the peak. In the transitional region   we 
expect a power-law slope, $\Omega_{gw} \propto f^\gamma$ with $\gamma$ varying between 1
and 3. For the model of parametric resonance that we considered in \cite{DBFKU}, we observed 
the $\Omega_{gw} \propto f$ behaviour.
 For tachyonic preheating, however, 
the source varies typically with a characteristic time given by $1/k_*$, i.e. 
more rapidly than the Green's function for any $k < k_*$. Therefore, we expect the 
IR tail $\Omega_{gw} \propto f^3$ to be valid quickly below the peak. For the 
IR part of the spectrum just below the peak that we can probe in our simulations, 
we found  $\gamma \sim 2 - 2.5$. The exact 
value was different for the three cases of significant initial velocity, negligible 
initial velocity and $g^2 << \lambda$, but it was the same in all our numerical results 
for a given case.

%%%%%%%%%%%%%%%%%%%%%%%%%%%%%%%%%%%%%%%%%%%%%%%%%%%%%%%%%%%%%%%%%%%%%%%%%%%%%%%%

\section{Numerical Results and Parameter Dependence}
\label{numerics}

\begin{figure}[htb]
\centering \leavevmode \epsfxsize=\columnwidth
\epsfbox{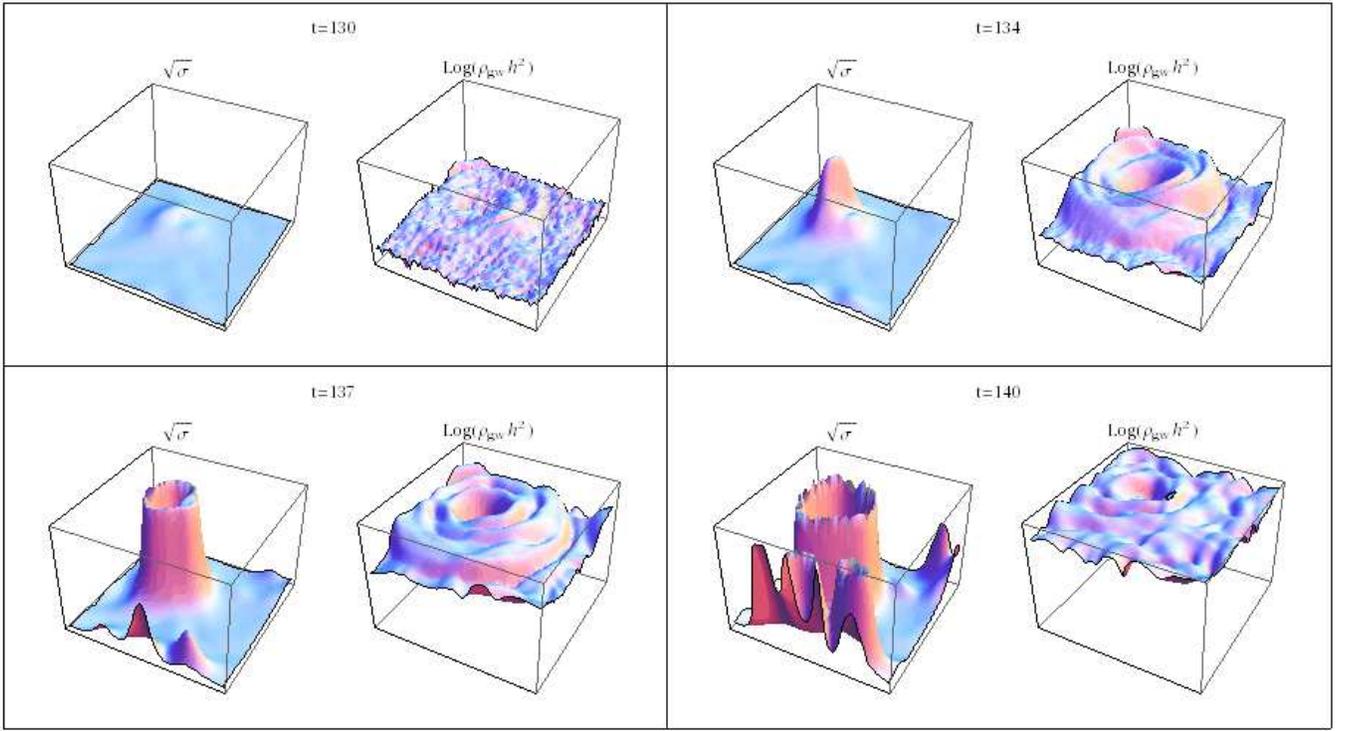}
\caption{Field amplitude and gravity wave density as a function of
  space on a two dimensional slice through the lattice. The slice is
  chosen at the height where the first Higgs bubble appeared. The
  simulation is of the model (\ref{hybrid}) with $\lambda=10^{-5}$,
  $\lambda/g^2=0.5$, $v=10^{-3}$, and $V_c=0$.}
\label{gwbubble}
\end{figure}

In this section we report the results of numerical calculations of GW
radiation from tachyonic preheating after hybrid inflation. Some
technical details of the numerical simulations are described in the
Appendices.  In particular, we study how the gravity wave
characteristics depend on the parameters of the hybrid model.  This
will allow us to directly check our analytical estimates and find fits
for the peak frequency and amplitude.  The analytical estimates of
Section \ref{analytics} were based on the picture of a sudden growth
of the density ``bubbles'' associated with the high peaks of the
random gaussian fields of initial quantum fluctuations.  Therefore we
begin this Section with an illustration of the growth of these
bubbles, shown in Fig.~\ref{gwbubble}.  The four panels show the field
amplitude (left of each panel) and GW energy density (right of each
panel) at four moments of time on a two-dimensional slice through the
three-dimensional lattice. The initial bump of the field produces an
exponentially growing bubble, and we can clearly see a burst of GW
radiating from this growing bubble. The background stochastic GW
radiation is the superposition of such concentric bursts.

The configuration space picture of Fig.~\ref{gwbubble} is complemented
by the momentum space spectra of gravity waves.
\begin{figure}[htb]
\centering \leavevmode \epsfxsize=8cm
\epsfbox{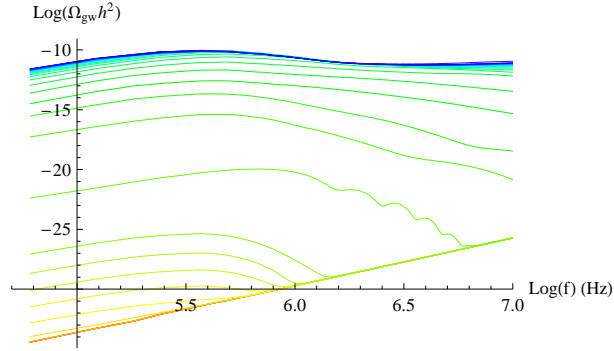}
\caption{Gravity wave spectra for the model (\ref{hybrid}) with
  $\lambda=10^{-14}$, $\lambda/g^2=0.5$, $v=10^{-3}$, and
  $V_c=10^{-5}$. Lower (red) curves correspond to earlier moments of the
  simulation and higher (bluer) curves  are from later moments  of the
  simulation. All results are scaled to the present-day spectrum.}
\label{l14V-5}
\end{figure}
Fig.~\ref{l14V-5} shows the gravity wave spectra from the lattice
simulation with one representative set of parameters corresponding to
the case $\lambda/g^2=1/2$ and the onset of preheating dominated by
classical rolling.  All spectra shown in this paper are scaled to
present-day units. The different curves show cumulative results for
different moments of times during the simulation. For some time after
inflation no gravity waves are produced, then there is a burst of
growth during preheating, and then the spectrum saturates at a
stationary level. A clear peak frequency is established early in the
growth and remains roughly constant as the amplitude grows.

As noted above, the model (\ref{hybrid}) involves essentially four
parameters related to preheating: the coupling constants $\lambda$ and
$g$, the symmetry breaking VEV $v$, and the unitless field velocity at
the critical point, $V_c$.

First, we discuss how GW spectra calculated from lattice simulations
depend on the parameters in the case where $\lambda = 2 g^2$. From the
point of view of numerical simulations this is the simplest case
because there is no gap between different mass scales.  Later we will
study how the results depend on the ratio $g^2 / \lambda$.

As noted in Section \ref{model}, the dependence of the spectra on
$v$ can be calculated analytically in the absence of
expansion. Specifically, changing $v$ does not shift the frequencies
at all and changes the density in gravity waves today proportionally
to $v^2$. This result should hold as long as $v$ is small enough to
maintain the waterfall condition.

As we anticipated in Section \ref{analytics}, variation with $V_c$,
the inflaton velocity at the bifurcation point, results in significant
variations of the GW spactra.  For the case with significant initial
velocity Eqs.~(\ref{fstarVc}), (\ref{OmegastarVc}) accurately predict
the peak frequency and amplitude to within an order of magnitude for
all of the parameters we have tested, which includes initial
velocities ranging from $V_c=10^{-5}$ to $10^{-2}$ for
$\lambda=10^{-14}$ and $10^{-5}$.

\begin{figure}[htb]
\begin{minipage}[t]{8cm}
\centering \leavevmode \epsfxsize=8cm
\epsfbox{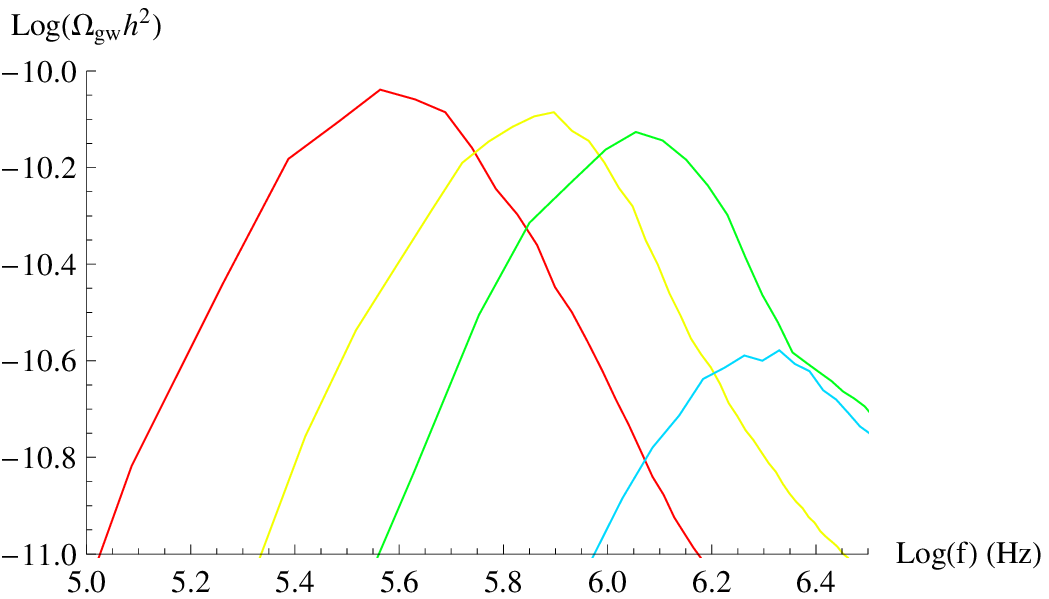}
\caption{Dependence of GW final spectra on $V_c$. The curves from left
  to right correspond to $V_c=10^{-5}$, $V_c=10^{-4}$, $V_c=10^{-3}$,
  and $V_c=10^{-2}$. Here $\lambda=10^{-14}$.  }
\label{l14Vdependence}
\end{minipage}
\hspace*{1.5cm}
\begin{minipage}[t]{8cm}
\centering \leavevmode \epsfxsize=8cm
\epsfbox{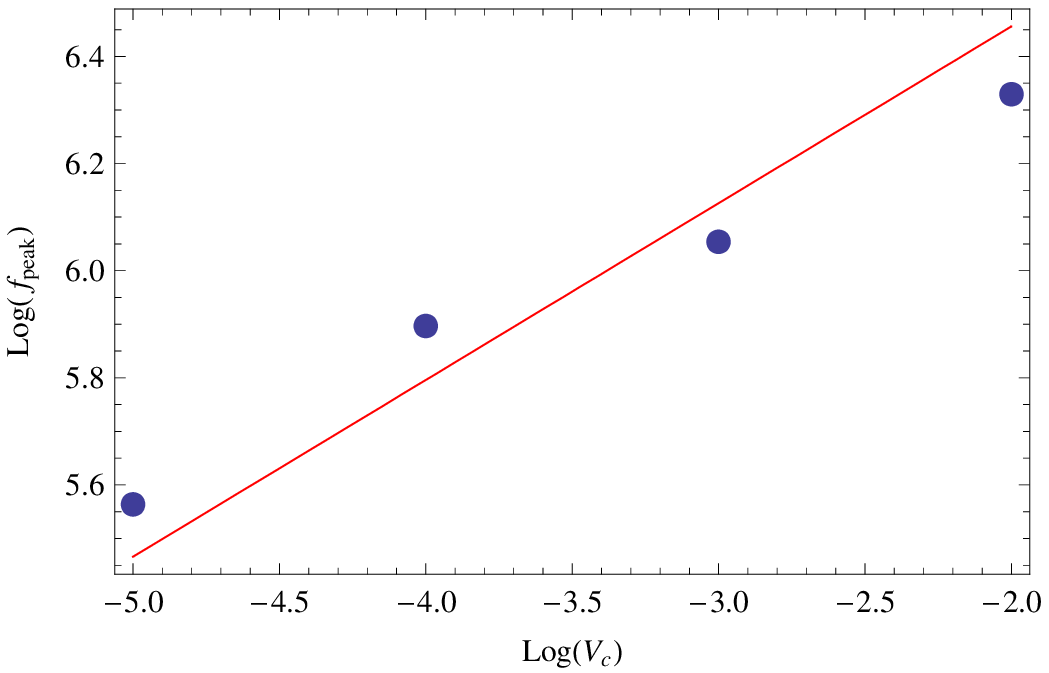}
\caption{The peak frequency for the gravity wave spectrum as a
  function of initial velocity for $\lambda=10^{-14}$. The line shows
  the slope of the prediction (\ref{fstarVc}) with the overall height
  adjusted to fit the data.}
\label{peakVdependence}
\end{minipage}
\end{figure}

Fig.~\ref{l14Vdependence} shows the final gravity wave spectra for
a range of initial velocities for $\lambda=10^{-14}$.  Increases in
$V_c$ lead to increases in peak frequency and decreases in peak
amplitude, but the shape of the spectra are quite similar. 
Fig.~\ref{peakVdependence} shows the numerical peak frequencies for the
cases from Fig.~\ref{l14Vdependence}. Eq.~(\ref{fstarVc}) predicts
that $f_{peak} \propto V_c^{1/3}$ so the figure shows a best-fit line
with slope $1/3$, illustrating that the dependence matches the
prediction. The actual values of the peak frequencies are all within a
factor of $3$ of the Eq.~(\ref{fstarVc}), which is in any case only an
order of magnitude estimate.

\begin{figure}[htb]
\begin{minipage}[t]{8cm}
\centering \leavevmode \epsfxsize=8cm
\epsfbox{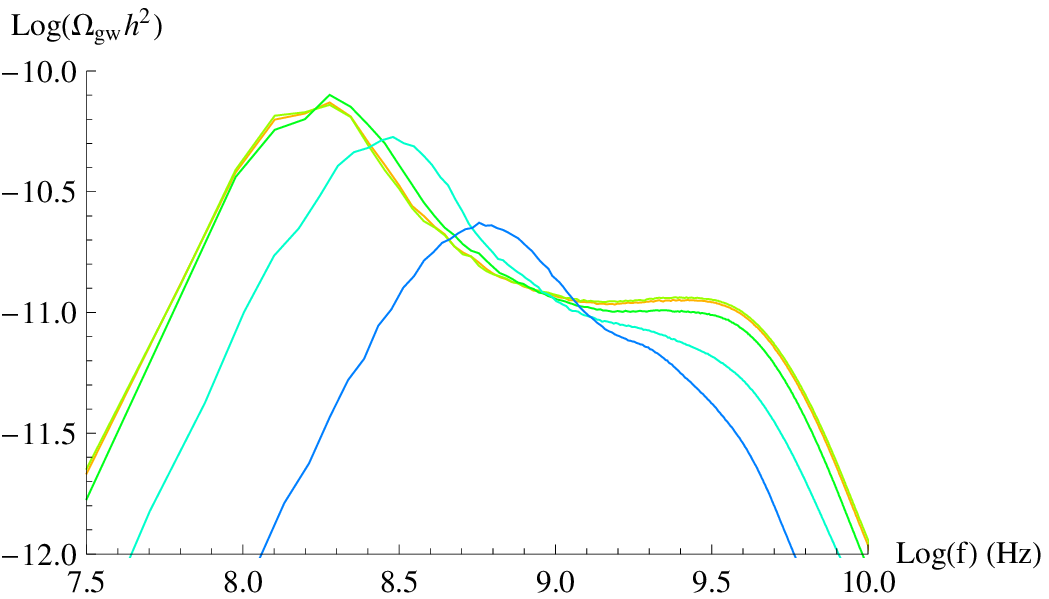}
\caption{Final gravity wave spectra after preheating for
  $\lambda=10^{-5}$. The plots from left to right show $V_c=: 0,
  10^{-6}, 10^{-5}, V_c=10^{-4}, 10^{-3}, 10^{-2}$. The plots for $V_c=0,
  10^{-6}$, and $10^{-5}$ lie almost perfectly on top of each other.}
\label{l5Vdependence}
\end{minipage}
\hspace*{1.5cm}
\begin{minipage}[t]{8cm}
\centering \leavevmode \epsfxsize=8cm
\epsfbox{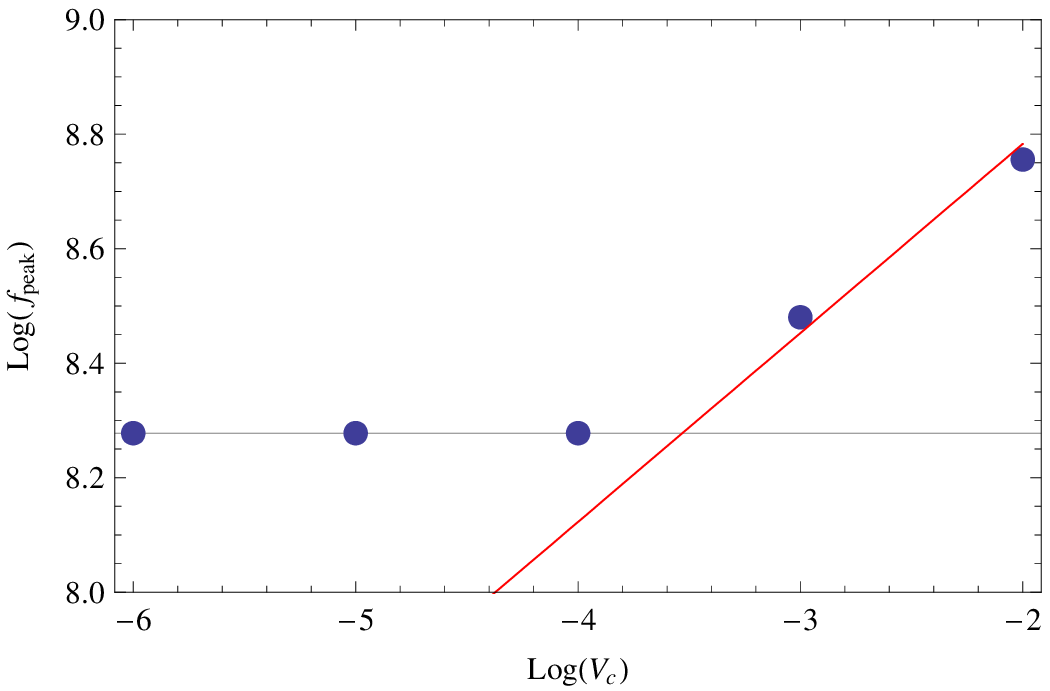}\\
\caption{The peak frequency for the gravity wave spectrum as a
  function of initial velocity for $\lambda=10^{-5}$. The line shows
  the slope of the prediction (\ref{fstarVc}) with the overall height
  adjusted to fit the last two points. The horizontal line is the peak
  frequency for $V_c=0$. For $V_c \lesssim 10^{-4}$ the peak remains
  constant at the $V_c=0$ value.}
\label{l5peakVdependence}
\end{minipage}
\end{figure}

Eq.~(\ref{fstarVc}) is not expected to be accurate in the limit of
very low initial velocities, as discussed in sub-section
\ref{lowvinit}. For example, Figs.~\ref{l5Vdependence}-\ref{l5peakVdependence} 
show the spectra and peak
frequencies for $\lambda=10^{-5}$. For this larger value of $\lambda$
the formula (\ref{fstarVc}) fails for $V_c \lesssim 10^{-4}$. Figure
\ref{l5Vdependence} shows that the spectra for $V_c=10^{-5}$ and
$V_c=10^{-6}$ are nearly identical to the $V_c=0$ spectrum, so below
$V_c \lesssim 10^{-4}$ there is no $V_c$ dependence. The
fit shown in this figure once again has the slope predicted by
Eq.~(\ref{fstarVc}) and a height adjusted to fit the data, but in this
case the fit is to the last two points only. The actual values for
peak frequency for these last two points match Eq.~(\ref{fstarVc}) to
within a factor of $2$.

For lower $\lambda$, such as the results shown above, the initial
velocity for which quantum diffusion dominates is much lower and thus
the results in Fig.~\ref{peakVdependence} show the $V_c^{1/3}$
dependence discussed above.

For $g^2 = 2\lambda$, our estimate (\ref{condVnegl}) indicates that
the velocity $V_c$ could be considered negligible below the cutoff
$V_c \approx C^3 \, \lambda^{3/2}$.  From the data above we conclude
that for $\lambda=10^{-5}$ this cutoff occurs between $V_c=10^{-5}$
and $V_c=10^{-4}$, which means $C \approx 10$. It is hard to estimate,
however, if this value is independent of $\lambda$. For lower values
of $\lambda$ we were unable to simulate slow enough initial velocities
to see the crossover between these two regimes. We can thus only
conjecture that this value $C \approx 10$ will be roughly constant for
different values of $\lambda$. This conjecture must be true if our
predicted dependence $f_{peak} \propto \lambda^{3/4}$ is correct for
the case of negligible initial velocity.

\begin{figure}[htb]
\centering \leavevmode \epsfxsize=8cm
\epsfbox{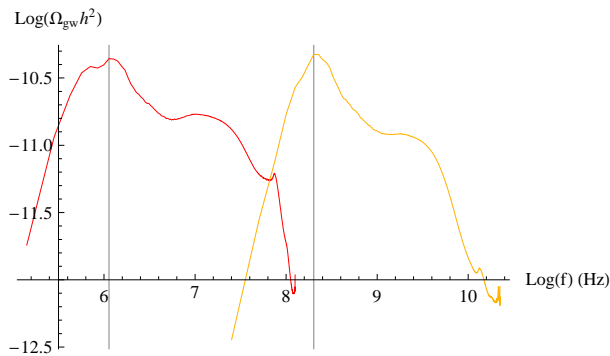}
\caption{Gravity wave spectra for $\lambda=10^{-14}$ (left) and
  $\lambda=10^{-5}$ (right), each with $V_c=10^{-3}$. The vertical line
  through the $\lambda=10^{-5}$ peak marks the maximum of that
  peak. The vertical line through the $\lambda=10^{-14}$ peak is a
  prediction based on extrapolating the $\lambda=10^{-5}$ peak
  frequency assuming $f_{peak} \propto \lambda^{1/4}$.}
\label{ldependence}
\end{figure}

Next, we investigate an impact of variation with $\lambda$ on the GW production.
For the case with significant initial velocity $V_c$  the equations above
predict that changing $\lambda$ should not change the amplitude but
should simply shift the frequency as $f_{peak} \propto \lambda^{1/4}$. 
Fig.~\ref{ldependence} clearly shows this to be the case.

As discussed above, we can not numerically test the dependence of the
spectrum on $\lambda$ for the case of negligible initial velocity, but
we believe our predictions should be accurate for this case because at
low initial velocity the bubble description of preheating is accurate.

We now discuss how our numerical results for the gravity wave spectrum depend
on the ratio $g^2 / \lambda\,$. For a given inflationary potential 
$V_{\mathrm{inf}}(\phi)$, changing $g^2 / \lambda$ changes the critical point 
$\phi_c$ and the unitless  velocity $V_c$ at that point, see (\ref{defVc}). 
However, in order to isolate the $g^2/\lambda$ dependence, we will keep $V_c$ 
constant in this section. We first consider the case of non-negligible initial 
velocity.

\begin{figure}[htb]
\centering \leavevmode \epsfxsize=11cm
\epsfbox{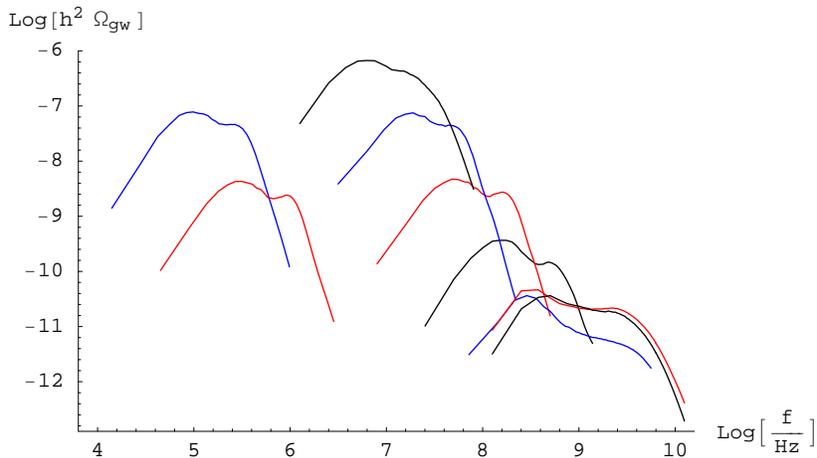}
\caption{Numerical results for the gravity wave spectra for several values 
of $\lambda / g^2$, for $\lambda=10^{-14}$ (left) and $\lambda=10^{-5}$ 
(right), each with $V_c = 10^{-3}$. The spectra for $\lambda=10^{-5}$ 
(right) are, from top to bottom, for $\lambda / g^2$ $=$ $20000$ (black), 
$5000$ (blue), $500$ (red), $50$ (black), $0.5$ (blue), $0.005$ (red) and 
$0.0005$ (black). 
The spectra for $\lambda=10^{-14}$ (left) are for $\lambda / g^2$ $=$ $5000$ 
(top, blue) and $500$ (bottom, red). To ease the comparison, all the 
spectra are shown for $v = 10^{-3}\,\Mp$, although for $g^2 << \lambda$ a 
lower value of $v$ may be necessary to consistently neglect expansion of the 
universe (see the main text for details). This lowers the spectra as 
$h^2\,\Omega_{\mathrm{gw}} \propto v^2$. In other words, the $y$-axis of 
the plot is really the log of $h^2\,\Omega_{\mathrm{gw}} / (10^3\,v / \Mp)^2\,$.}
\label{omegalg}
\end{figure}

In this case, gravity wave spectra for different values of $g^2 /
\lambda$ are shown in Fig.~\ref{omegalg}, for $V_c = 10^{-3}$. For
$g^2 \gtrsim \lambda$, the peak frequency and amplitude are roughly
independent of $g^2 / \lambda$, in agreement with our original
prediction, Eqs.~(\ref{fstarVc}), (\ref{OmegastarVc}).  (When $g^2$
increases, rescattering effects become more important and the UV part
of the GW spectrum has higher amplitude). On the other hand, for $g^2
<< \lambda$, it is clear from Fig.~\ref{omegalg} that decreasing $g^2
/ \lambda$ decreases the peak frequency and increases the peak
amplitude, in aggreement with
Eqs.~(\ref{fstarg}), (\ref{Omegastarg}). Note also that, for a given
value of $g^2 / \lambda << 1$, the dependence on $\lambda$ is very
well given by $f_* \propto \lambda^{1/4}$ and
$h^2\,\Omega^*_{\mathrm{gw}}$ independent of $\lambda$, again in
agreement with Eqs.~(\ref{fstarg}), (\ref{Omegastarg}).  Note that
Fig.~\ref{omegalg} shows the gravity wave spectrum for $v =
10^{-3}\,\Mp$. For $g^2 << \lambda\,$, a lower value of $v$ may be
necessary to consistently neglect the expansion of the universe, as
discussed below. Lowering $v$ leaves the shape of the spectra
unchanged but reduces their amplitude proportionally to $v^2$.

\begin{figure}[hbt]
\begin{center}
\begin{tabular}{cc}
\centering \leavevmode \epsfxsize=7cm
\epsfbox{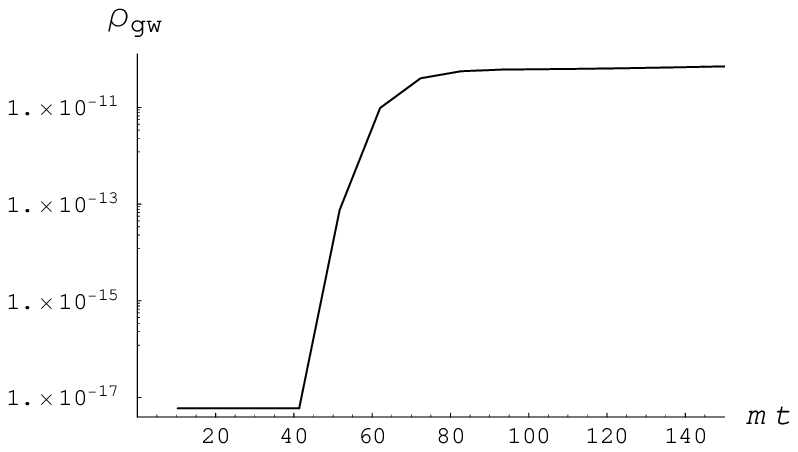} &
\centering \leavevmode \epsfxsize=7cm
\epsfbox{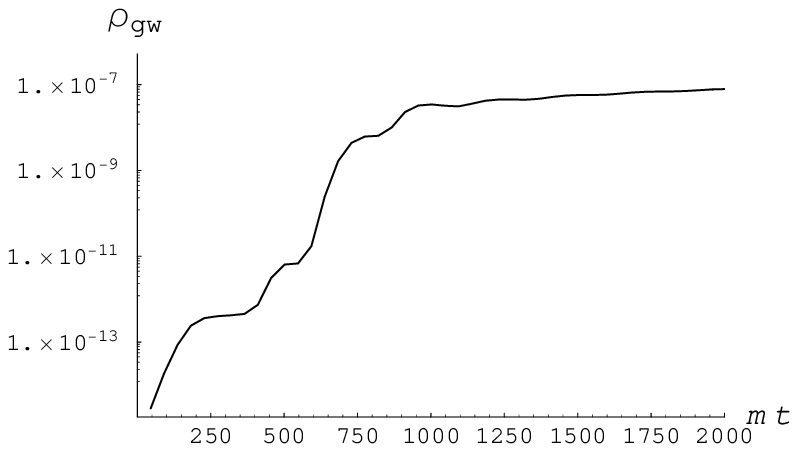}
\end{tabular}
\caption{Evolution with time of the total energy density in gravity waves, 
$\rho_{\mathrm{gw}} / (10^3\,v / \Mp)^2\,$, for $\lambda / g^2 = 0.5$ 
(left) and $\lambda / g^2 = 2000$. The parameters are the same as in 
Fig. \ref{meanslg}.}
\label{rhogwlg}
\end{center}
\end{figure}

Decreasing $g^2/\lambda$ results in qualitatively different GW
production.  Fig.~\ref{rhogwlg} shows the accumulation with time of
the total energy density in gravity waves for two cases, the left
panel for $g^2 \sim \lambda$ and the right panel for $g^2 << \lambda$.
In the first case we see a single burst of GW production from
tachyonic preheating.  In the second case, we clearly see successive
bursts of gravity wave production, due to the successive tachyonic and
non-adiabatic amplifications discussed in subsection \ref{g2lll}, and
the subsequent bubble collisions. Because the characteristic time
scale in this case is given by $m\,\sqrt{\lambda} / g\,$, for $g^2 <<
\lambda$ preheating and gravity wave production take more time than
for $g^2 \sim \lambda$. As a result, the upper bound on $v$ necessary
for the expansion of the universe to be negligible is lower than for
$g^2 \gtrsim \lambda$. This decreases the maximum GW amplitude as
$h^2\,\Omega_{\mathrm{gw}} \propto v^2$. If the peak of the gravity
wave spectrum is reached after a proper time $\tau\,$, we require it
to be much smaller than the Hubble time \be
\label{negla}
\frac{\tau}{H^{-1}} \,\approx\, m\,\tau\;\frac{v}{\Mp} << 1 \, .
\ee

In fact, for all the cases with $g^2 << \lambda$ that we considered,
the peak amplitude of the GW spectrum continued to slightly but
constantly increase with time after the last burst of GW production
(i.e. after $m\,t \sim 1000$ for the right panel of
Fig.~\ref{rhogwlg}), and the peak frequency tended to move further
towards the infrared. However, if we consider very late times
expansion will become significant, thus invalidating the results of
our simulations. For low values of $v$ this growth could continue
longer without being diluted by expansion, but the overall amplitude
would be lower. For each set of parameters $\lambda$, $g$, and $V_c$
there is thus an optimal value of $v$ that will produce the greatest
amplitude of gravity waves before expansion becomes significant. For
each set of parameters $\lambda$, $g$, and $V_c$ we define $v_{opt}$
to be this optimal value and we define the time $\tau$ as the time at
which expansion would become significant for $v=v_{opt}$. The spectra
shown in Fig.~\ref{omegalg} were obtained at that time.

\begin{figure}[hbt]
\begin{center}
\begin{tabular}{ccc}
\centering \leavevmode \epsfxsize=5.9cm
\epsfbox{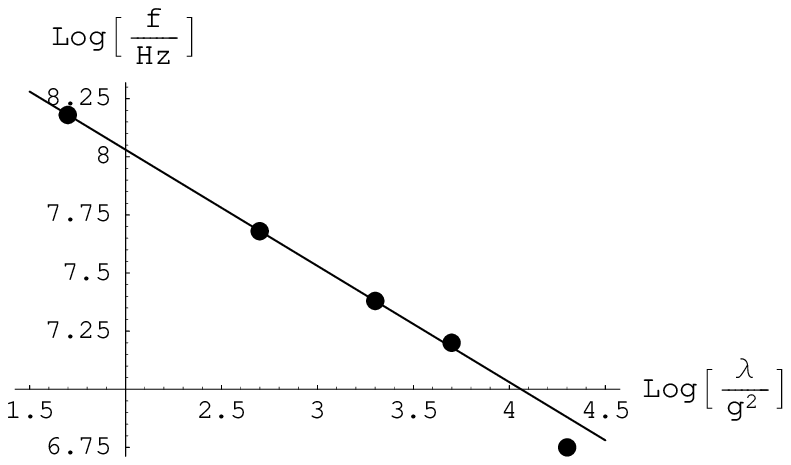} &
\centering \leavevmode \epsfxsize=6.6cm
\epsfbox{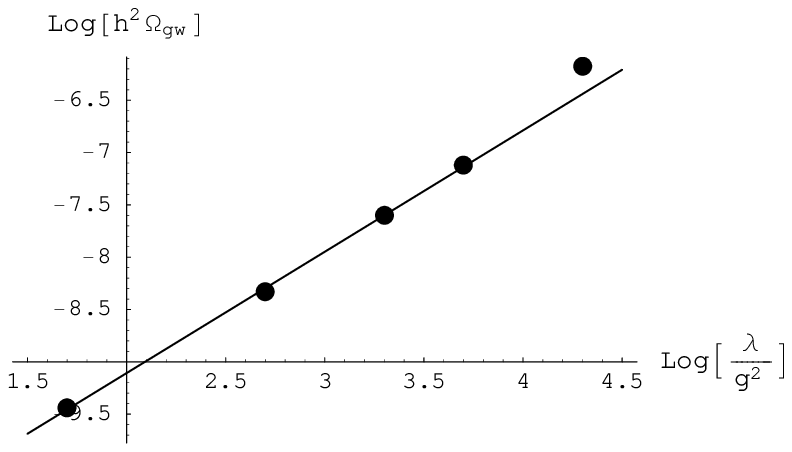} &
\centering \leavevmode \epsfxsize=5.9cm
\epsfbox{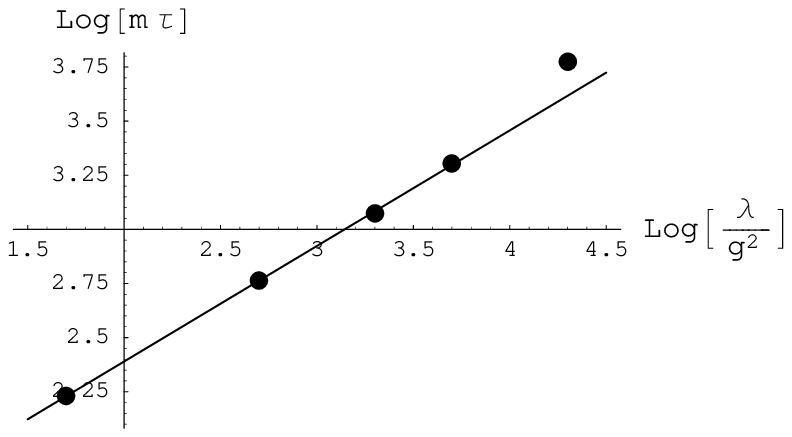}
\end{tabular}
\caption{Peak frequency $f_*$ (left), peak amplitude
  $h^2\,\Omega^*_{\mathrm{gw}}$ (middle) and time $m\,\tau$ defined in
  the main text (right), for $V_c = 10^{-3}$, $\lambda = 10^{-5}$,
  $v=10^{-3}$, and several values of $g^2 / \lambda << 1$. The lines
  shown are the best linear fits for all the points except the one
  with $\lambda/g^2 = 20000$ (extreme right): $f_* \propto
  (\lambda/g^2)^{-1/2}$, $h^2 \Omega^*_{\mathrm{gw}} \propto
  (\lambda/g^2)^{1.16}$ and $m \tau \propto
  (\lambda/g^2)^{0.53}\,$. }
\label{fitslg}
\end{center}
\end{figure}

Fig.~\ref{fitslg} shows in the left panel: the peak frequency, in the
middle: the peak amplitude and in the right: the time $m\,\tau$
defined above, for $V_c = 10^{-3}$, $\lambda = 10^{-5}$ and several
values of $g^2 / \lambda << 1\,$. We see that the prediction
(\ref{fstarg}), $f_* \propto g$, is very well satisfied. The peak
amplitude grows slightly more rapidly than (\ref{Omegastarg}) when
$g^2 / \lambda$ decreases, $h^2\,\Omega^*_{\mathrm{gw}} \propto
(\lambda/g^2)^{1.16}\,$. Finally, $m\,\tau$ grows slightly more
rapidly than $\sqrt{\lambda}/g\,$, $m \tau \propto
(\lambda/g^2)^{0.53}\,$. In these fits, we disregarded the point at
the extreme right ($\lambda / g^2 = 20000$), which is slightly
displaced compared to the others. In fact, for $\lambda / g^2 \gtrsim
10^4\,$, and for the initial velocity $V_c = 10^{-3}$ considered here,
the inflaton reached $|\phi| > \phi_c$ and spent most of the time in
these regions. Indeed, in these regions, the minimum in the
$\sigma$-direction is at $\sigma = 0$ and the effective mass for
$\phi$ is very low, $m^2_\phi = g^2 \,
\langle\sigma^2\rangle$. Apparently, this slightly increases the peak
amplitude and decreases the peak frequency. However, it may not be a
good approximation to neglect $V_{inf}(\phi)$ in the region where
$|\phi| > \phi_c\,$.  Also, we decreased $g^2 / \lambda$ while keeping
$V_c$ fixed. However, for a given physical initial velocity
$\dot{\phi}_c\,$, decreasing $g^2/\lambda$ decreases the unitless
velocity $V_c\,$, see (\ref{defVc}). For sufficiently small $V_c\,$,
the inflaton will not reach values $|\phi| > \phi_c\,$. We will thus
neglect this effect below, although it may be useful to further
decrease the peak frequency.

We then obtain the following fits for the peak frequency and amplitude of the GW 
spectrum for the case  $g^2 / \lambda << 1$
\bea
\label{fstargnum}
f_* \simeq \frac{g}{\sqrt{\lambda}}\,\lambda^{1/4}\,10^{10.25}\;\mathrm{Hz} \\
\label{Omegastargnum}
h^2\,\Omega^*_{\mathrm{gw}} \simeq 10^{-5.5}\;\left(\frac{\lambda}{g^2}\right)^{1.16}
\;\left(\frac{v}{\Mp}\right)^2 \, .
\eea
From the fit of $m\,\tau$, the constraint (\ref{negla}) for the expansion of the universe 
to be negligible becomes
\be
\label{neglanum}
\frac{v}{\Mp} << 10^{-1.3}\;\left(\frac{\lambda}{g^2}\right)^{-0.53} \, .
\ee
Note that this condition always implies $\phi_c << \Mp\,$. Together with 
(\ref{Omegastargnum}), it also implies 
$h^2\,\Omega^*_{\mathrm{gw}} << 10^{-8.1}\,(\lambda / g^2)^{0.1}\,$. 

For instance, for $\lambda \sim 0.1$, $f_* \lesssim 10^3\,\mathrm{Hz}$
for $g^2 \lesssim 10^{-15}\,$, and the amplitude (\ref{Omegastargnum})
may be relevant for Advanced LIGO while satisfying the constraint
(\ref{neglanum}).

We expect similar results as above in the case of negligible velocity
$V_c$. In this case, where quantum diffusion dominates,
observationally more interesting results can be obtained without the
requiremenet $g^2 << \lambda$.

\section{Discussion and Perspectives}
\label{Discussion}

In this paper, we studied the stochastic background of gravitational waves produced from 
tachonic preheating after hybrid inflation in the very early universe. The present-day 
frequencies and amplitudes of these gravity waves may cover a wide range of values, depending 
on the main phenomenological parameters introduced in Section II: the VeV $v$ and self-coupling 
$\lambda$ of the symmetry breaking fields, their coupling to the inflaton $g^2$ and the (unitless) 
initial velocity of the inflaton at the critical point $V_c$. We developed analytical and numerical 
tools to calculate the resulting GW spectra and to study in detail how they depend on these 
parameters. We identified three dynamical regimes leading to qualitatively different results for 
the GW spectra produced from tachyonic preheating: (i) the case with $g^2 \gtrsim \lambda$ and the 
onset of preheating driven by the classical rolling of the inflaton, (ii) the case with 
$g^2 \gtrsim \lambda$ and quantum diffusion at the onset of preheating (corresponding to a negligible 
initial velocity of the inflaton at the bifurcation point) and (iii) the case with $g^2 << \lambda$ 
where GW are produced in successive bursts. Only the first case was considered in previous works 
about GW from tachyonic preheating \cite{juan1, juan2}, where the resulting GW spectrum was computed 
in the specific range $g^2 \sim \lambda \sim V_c \sim \mathcal{O}(1)$.

Based on our results, we can determine the range of parameters of hybrid inflation models for which 
preheating may lead to a GW signal that is potentially observable. Let us first discuss the frequencies 
and amplitudes of stochastic GW backgrounds that are relevant for GW astronomy.
\begin{figure}[htb]
\centering \leavevmode \epsfxsize=18cm
\epsfbox{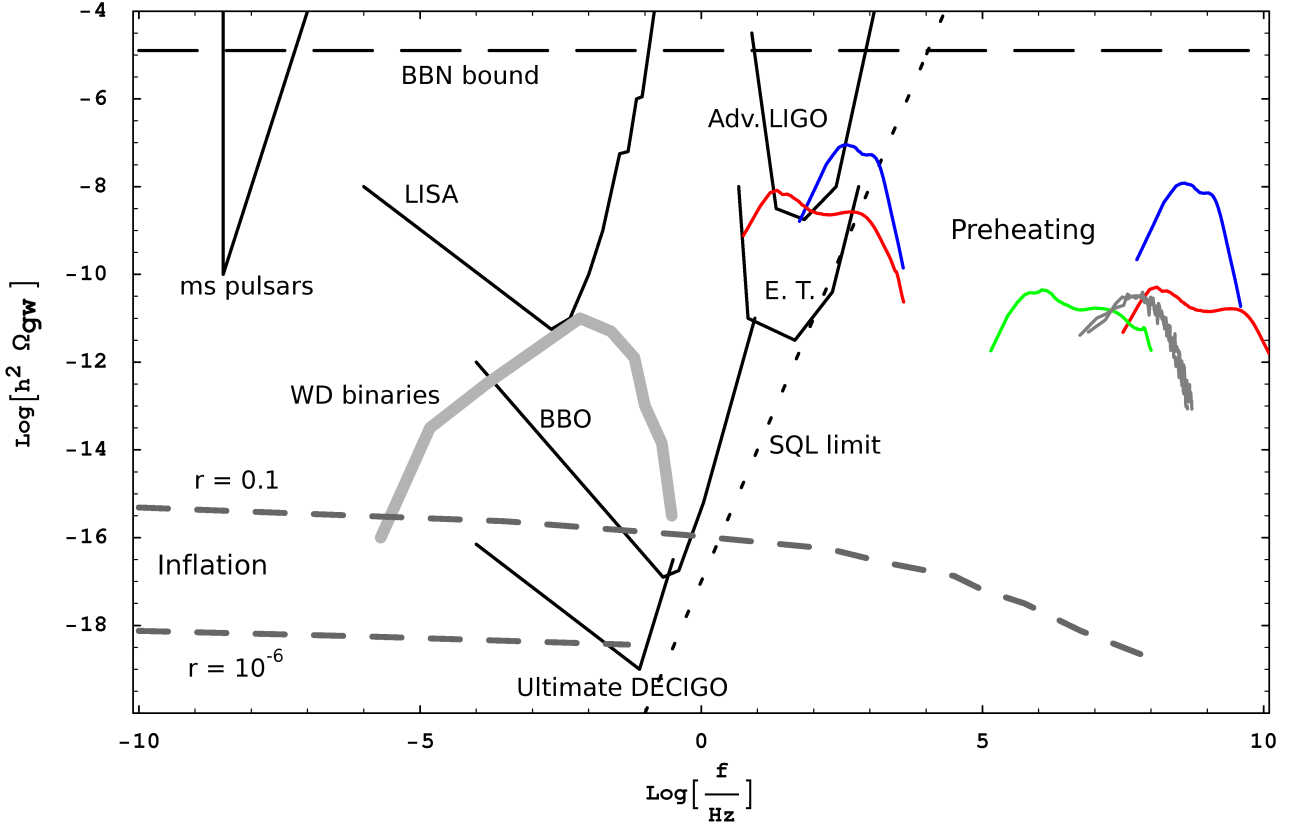}
\caption{Expected sensitivities of interferometric experiments (adapted from \cite{buonanno}) compared to gravity 
wave spectra from tachyonic preheating in the cases of significant intial velocity with $g^2 \gtrsim \lambda$ 
(green), negligible initial velocity with $g^2 \gtrsim \lambda$ (red), and $g^2 << \lambda$ (blue). Also shown 
are the BBN and ms pulsar bounds, the stochastic background from extragalactic White Dwarfs binaries (taken 
from \cite{phinney}), the inflationary background for two values of the tensor to scalar ratio, and the spectrum 
from preheating in a $\lambda \, \phi^4$ model of chaotic inflation (double spectra in grey at the right, calculated 
with our previous code \cite{DBFKU}). The GW spectra from tachyonic preheating, for the peak from left to right, 
correspond to the following values of the parameters:
(1) $\lambda = 2 g^2 = 10^{-14}$, $v = 3\,10^{-7}\,\Mp$ and negligible initial velocity ; 
(2) $\lambda = 0.1$, $g^2 = 10^{-16}$ and $v = 3\,10^{-10}\,\Mp$ (independent of $V_c$) ; 
(3) $\lambda = 2 g^2 = 10^{-14}$, $v = 10^{-3}\,\Mp$ and $V_c = 10^{-3}$ ; 
(4) $\lambda = 0.1$, $g^2 = 10^{-4}$ and $v = 6\,10^{-4}\,\Mp$ (independent of $V_c$) ; 
(5) $\lambda = 2 g^2 = 10^{-5}$, $v = 10^{-3}\,\Mp$ and negligible initial velocity.}
\label{sensi}
\end{figure}
We consider the present-day frequencies $f$ and the spectrum of energy density per logarithmic frequency 
interval $h^2\,\Omega_{\mathrm{gw}}(f)$. Fig.~\ref{sensi} shows the sketch of expected sensitivities of planned 
and future interferometric experiments: Advanced LIGO, LISA, the Einstein Telescope, Big Bang Observer and 
DECIGO (see \cite{buonanno} for the relevant references). Note also the recent atomic interferometric sensor 
proposal of \cite{agis}, which may be sensitive to GW backgrounds with frequencies in the range $1-10$ Hz 
and in the LISA range. We also show the BBN and ms pulsar bounds, and predictions for the stochastic GW 
background generated from inflation, for different values of the parameter $r$ (the ratio of the tensor and scalar 
amplitudes of the inflationary cosmological perturbations). Cosmological GW signals will be obscured by several 
expected astrophysical foregrounds, notably from White Dwarf binaries \cite{phinney}, see Fig.~\ref{sensi}.

There are also several bar and spherical resonant detectors (see
e.g. \cite{minigrail}) operating in the kHz range. Other experiments
have been proposed at higher frequencies, up to $100$ MHz
\cite{Nishizawa:2007tn}. However, the sensitivity to
$h^2\,\Omega_{\mathrm{gw}}$ \footnote{The concept of
  $h^2\,\Omega_{\mathrm{gw}}$ for stochastic backgrounds is not
  applicable for GW signals peaked within a very narrow frequency
  band.} drops dramatically when the frequency increases. Indeed, we
also have to take into account the current ``Standard Quantum Limit'' 
\begin{equation}\label{SQL}
h^2\,\Omega_{\mathrm{gw}} \sim \frac{10^{-17}}{x^2}  \, \left(\frac{f}{Hz}\right)^3 \ ,
\end{equation}
shown in Fig.~\ref{sensi}. The region on the right of this line is not observable with interferometric 
experiments due to the shot noise fluctuations of photons. Here $x$ is the squeezing parameter, which 
experimentalists are trying to push from one to a few.

Exemples of GW spectra from tachonic preheating calculated in this paper are shown in Fig.~\ref{sensi}. 
We also show in this figure a GW spectrum from preheating after chaotic inflation calculated in 
\cite{DBFKU} (the double-spectrum in grey), which is located in the high-frequency side and is not 
observable. GW from tachyonic preheating may cover a much wider range of frequencies, spreading from 
the highest frequencies shown in the figure to the region in between the SQL limit and the WD binaries, 
which is in principle observable \footnote{For the  sake of curiosity, one can notice the analogy with 
the $2.73$K CMB radiation which is observable in the strip of frequencies between the borders of the 
foreground dust and synchrotron radiations, as was outlined in early 1960.}. In each case shown 
in the figure, a lower Higgs VeV would lead to lower amplitudes and smaller coupling constants would lead 
to smaller frequencies. 

\begin{figure}[hbt]
\centering \leavevmode \epsfxsize=18cm
\epsfbox{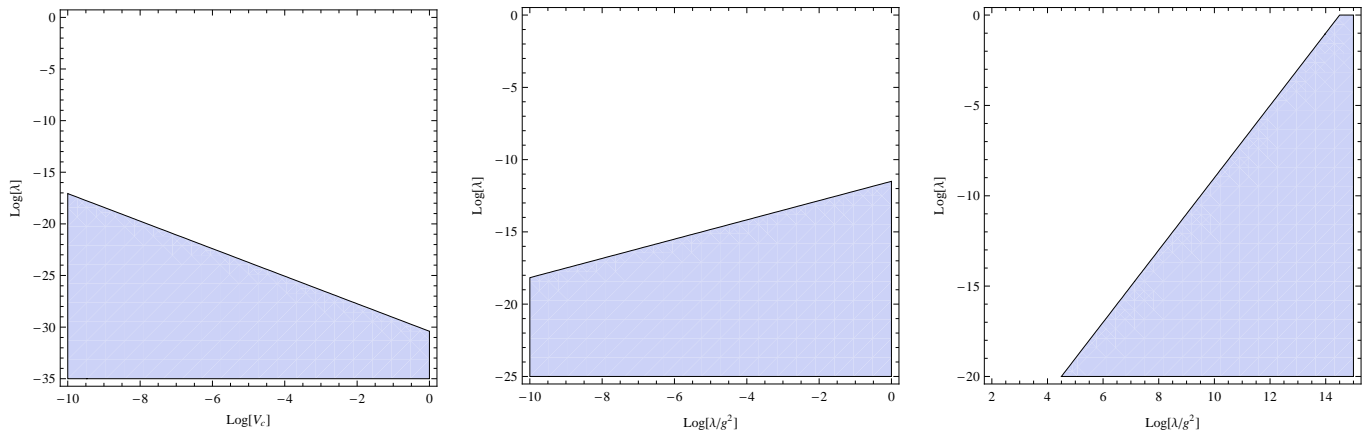}
\caption{
The regions of parameter space for which the peak of the
gravity wave signal satisfies $f_* < 10^3$ Hz. The panels
show: Left - significant initial velocity and $g^2 \gtrsim \lambda$. 
Middle - negligible initial velocity and $g^2 \gtrsim \lambda$. 
Right - $g^2 \ll \lambda$.}
\label{constraints}
\end{figure}

Our analytical and numerical results for the peak frequency $f_*$ and the peak amplitude $h^2\,\Omega_{\mathrm{gw}}^*$ 
of the GW spectra are well described by Eqs.~(\ref{fstarVc})-(\ref{OmegastarVc}) 
in the case with $g^2 \gtrsim \lambda$ and the onset of preheating driven by the classical rolling of the inflaton, 
Eqs.~(\ref{fstarV0})-(\ref{OmegastarV0}) in the case with $g^2 \gtrsim \lambda$ and 
quantum diffusion at the onset of preheating (negligible initial velocity of the inflaton at the 
critical point) and Eqs.~(\ref{fstargnum})-(\ref{Omegastargnum}) in the case with $g^2 << \lambda$.
The peak frequency depends essentially on the coupling constants and is independent of the Higgs VeV $v$. 
As a result, $\lambda \sim g^2 \sim \mathcal{O}(1)$ leads to GW at very high frequencies, independent 
of the energy scale during inflation, and the only way to lower the frequency is to lower the coupling 
constants. On the other hand, $h^2\,\Omega_{\mathrm{gw}}^* \propto v^2$, so that at any frequency the 
amplitude can be relatively high for sufficiently high $v$, roughly up to the maximal bound~\footnote{This bound 
follows from the approximate equation 
(\ref{omegaR}) with the constraint $k_* > H$ (meaning that the scalar fields and the gravity waves 
are produced inside the Hubble radius). The same constraint, $k_* > H$, is also roughly the condition 
for the expansion of the universe to be negligible and for the waterfall condition to be 
satisfied. It also implies that $v << \Mp$ and (for the small coupling constant 
we consider) that the inflationary model is not ruled out by the non-observation of 
inflationary gravity waves.} $h^2 \Omega_{\mathrm{gw}}^* < 10^{-6}\,$. 
Such a high energy density in GW implies in particular that these GW may 
already be observable by Advanced LIGO, but this generally requires very small coupling constant(s), 
see Fig.~\ref{sensi}. For illustration, we show in Fig.~\ref{constraints}, for each of the three cases 
discussed above, the range of parameters of the hybrid inflation models such that the peak frequency of the 
GW produced from preheating satisfy $f_* < 10^3$ Hz. This is basically the condition for these gravity 
waves to be observable if $v$ is set to the maximum possible value consistent with the waterfall condition.
In the first case of Fig.~\ref{constraints}, corresponding to $g^2 \gtrsim \lambda$ and significant initial 
velocity, the coupling of interest $\lambda$ typically has to be extremely small, down to $\lambda < 10^{-30}$ 
for $V_c \sim 1$. In the second case, with negligible initial velocity, we may have 
$g^2 \sim \lambda \sim 10^{-11}$. In the third case ($g^2 << \lambda$), the most interesting regime 
corresponds to $\lambda \sim 1$, but this still requires $g^2 < 10^{-14}$.

Above we focused on the peak frequency $f_*$ of the GW spectra. At lower frequencies, the spectra 
have a power-law tail, $\Omega_{\mathrm{gw}} \propto f^{\alpha}$. For tachyonic preheating, we found 
in this paper an intermediate regime with a relatively steep slope, $\alpha = 2 - 2.5$, before the IR 
tail with $\alpha = 3$ corresponding to the modes produced outside the Hubble radius. The infra-red part 
of the spectrum does not significantly improve the constraints on the parameters shown in 
Fig. \ref{constraints}. As outlined above, the range of parameters leading to the maximum amplitude 
in GW correspond to $k_*$ slightly below $H$. In this case, the spectra decrease as 
$\Omega_{\mathrm{gw}} \propto f^3$ for frequencies just below the peak. In particular, since this is 
the same dependence as the SQL limit (\ref{SQL}), if the peak of the GW spectrum is located at the right 
of this line, the same is true for the IR tail.

In sum, our sober conclusion is that GW from tachyonic preheating after hybrid inflation can fall 
in the observable range of the $h^2 \Omega_{\mathrm{gw}} -- f$ plane that is currently conceivable only 
for uncomfortably small values of the coupling constants of the model. 

We should note, however, that particle physics models of hybrid inflation with such small coupling constants have been considered. 
First, if the field $\sigma$ in the model (\ref{hybrid}) corresponds to a flat direction, we may expect its negative mass squared 
at the origin to be of the order $\lambda v^2 \sim \mathrm{TeV}^2$, and the true minimum of the potential to be located at a VeV 
at a much higher scale, as high as $v \sim M_\mathrm{Pl}$. This gives a coupling constant as low as $\lambda \sim 10^{-30}$. 
Hybrid inflation models along this line have been proposed in \cite{randall, stewart}. 
Another model which, for the same reason, naturally involves very small coupling constants is thermal inflation \cite{thermal}. 
Preheating mechanisms in this model have not been studied in detail yet, but tachyonic amplifications may lead to an interesting 
GW signal, see also \cite{wanil}. Another model of hybrid inflation in supergravity is P-term inflation \cite{pterm}, 
which includes F-term and D-term models as special cases and may also be realized in D3/D7 models in string theory. 
Satisfying the observational bounds on cosmic strings in P-term inflation may require small couplings 
($\lambda \sim g^2 < 10^{-10}$), see \cite{pterm}. Finally, we note that the case $g^2 << \lambda$ may have connections
with brane-anti-brane inflation in a warped throat \cite{braninf}, which is another prototype of hybrid inflation 
in string theory. According to \cite{neil}, the end of inflation in this 
scenario should exhibit similarities with the model (\ref{hybrid}) with $g^2 << \lambda$ (see in particular 
Eqs.~(30)-(32) of the second paper in \cite{neil}). In this case, a very small $g^2$ would emerge naturally 
because of the exponential warping of the throat. 

More generally, the hybrid inflation model (\ref{hybrid}) that we considered in this paper is only one 
of the much broader class of inflationary models where preheating occurs through tachyonic effects. GW 
production from other models involving tachyonic preheating is a subject for further investigations.

\vskip 0.5in
\vbox{
\noindent{ {\bf Acknowledgments} } \\ 
\noindent
We thank Juan Garcia-Bellido, Daniel Figueroa and Alessandra Buonanno for useful
discussions.  We also acknowledge the use of the CITA and IFT
computation clusters, and thank CITA and KITP for hosting
G.F. during part of this work. The work of J.F.D. was supported by the
Spanish MEC, via FPA2006-05807 and FPA2006-05423. G.F. was supported
by PHY-0456631. L.K. was supported by NSERC and CIFAR.}
\vskip 0.5in

\section*{Appendix A: Numerical Calculations}

Because of the range of scales needed to accurately determine the
spectra for our runs the simulations shown here were done using
CLUSTEREASY \cite{clustereasy}, the parallel programming version of
LATTICEEASY \cite{latticeeasy}. We evolve the non-$TT$ metric
perturbations along with the scalar fields using Eq.~(\ref{eom})
and then extract the $TT$ part and calculate the gravity wave spectrum
as describe in Section \ref{gwcalculations}.

For runs with significant initial velocity, we did some runs with the
initial conditions described in \cite{symbreak}, but we found the results 
virtually identical to those obtained by simply starting with vacuum initial 
conditions at the critical point, so the simulations shown here were all done 
starting from vacuum modes.

We found that the gravity wave spectrum is highly sensitive to the UV
and IR tails. When the lattice has insufficient UV coverage this
results in a large, spurious growth in the UV part of the
spectrum. This growth can in turn raise or lower the rest of the
spectrum, leading to a generally unreliable spectrum. Likewise, having
insufficient IR, even well below the region of the peak, can lead to
inaccurate results in the region of the peak. These effects are for
the most part not a result of the initial spectrum. For most of the
physical parameters described here we did runs with a wide range of
cutoffs in the initial spectrum and found the final results virtually
independent of these cutoffs. In short an accurate spectrum simply
required a grid with enough modes in a wide band around the peak.

\section*{Appendix B: Symmetries of the Metric Perturbations}

In principle the energy density in gravity waves (\ref{energy3}) involves 
a sum over the nine components of $h_{ij}$, but this can be
reduced through symmetry. The tensor $h_{ij}$ is symmetric
($h_{ij}=h_{ji}$, 3 equations), traceless ($h_{ii}=0$, 1 equation),
and transverse ($k_i h_{ij}=0$, 3 equations), which reduces the number
of independent components from nine to two. However, the two
components that you need to find the rest are different for different
$\hat{k}$ because some of components are zero along certain directions
in $k$ space. The scheme we used for calculating $\Omega_{gw}$ for all
directions in $k$ is the following:

For all points with $k_z \ne 0$ (off the $k_x$, $k_y$ plane) calculate
$h_{11}$ and $h_{12}$ and the symmetry equations give
\begin{equation}
h_{ij} h_{ij}^* = 2 {k_1^2 + k_2^2 + k_3^2 \over k_3^2 \left(k_2^2 +
  k_3^2\right)} \left[\left(k_1^2 + k_3^2\right) h_{11,R}^2 + 2 k_1
  k_2 h_{11,R} h_{12,R} + \left(k_2^2 + k_3^2\right) h_{12,R}^2\right]
  + \mbox{equivalent terms for $h_{ij,I}$}
\end{equation}

For all points with $k_z=0$ and $k_y \ne 0$ (on the $k_x$, $k_y$ plane
but off the $k_x$ axis) calculate $h_{11}$ and $h_{13}$ and the
corresponding solution is
\begin{equation}
2 {k_1^2 + k_2^2 \over k_2^4} \left[\left(k_1^2 + k_2^2\right)
  h_{11,R}^2 + k_2^2 h_{13,R}^2\right] + \mbox{equivalent terms for
  $h_{ij,I}$}
\end{equation}

For all points with $k_y=k_z=0$ (the $k_x$ axis) calculate
$h_{22}$ and $h_{23}$ and the corresponding solution is
\begin{equation}
2 \left(h_{22,R}^2 + h_{23,R}^2\right) + \mbox{equivalent terms for
$h_{ij,I}$}
\end{equation}

Analogous equations are used to calculation $h_{ij}' h_{ij}'^*$.

%%%%%%%%%%%%%%%%%%%%%%%%%%%%%%%%%%%%%%%%%%%%%%%%%%%%%%%%%%%%%

%%%%%%%%%%%%%%%%%%%%%%%%%%%%%%%%%%%%%%%%%%%%%%%%%%%%%%%%%%%%%%%%%%%%%%%%%

\end{document}